\documentclass[
floatfix,
twocolumn,
aps,
pra,
superscriptaddress,
10pt,
]
{revtex4-1}

\usepackage{tikz}
\usepackage{graphicx}
\usepackage{epstopdf}
\usepackage{braket}
\usepackage{amsmath, amsthm, amssymb}

\newcommand{\ben}{\begin{equation}}
\newcommand{\een}{\end{equation}}

\newcommand{\bea}{\begin{eqnarray}}
\newcommand{\eea}{\end{eqnarray}}
\def\sss{\scriptscriptstyle\rm}
\def\x{_{\sss X}}

\def\s{_{\sss S}}
\def\xc{_{\sss XC}}
\def\H{_{\sss H}}
\def\Hx{_{\sss HX}}
\def\Hxc{_{\sss HXC}}

\def\ext{_{\rm ext}}

\def\br{{\bf r}}

\def\mT{\mathcal{T}}

\def\hH{\hat{H}}

\def\hn{{\hat{n}}}

\def\1var{(\bx_1...\bx\N)}

\def\d{\delta}
\def\e{\epsilon}
 
\def\w{\omega}

\renewcommand{\Re}{\operatorname{Re}}

\newcommand*\diff{\mathop{}\!\mathrm{d}}
\def\dr{\diff\br}
\def\dt{\diff t}

\def\dr{\diff\br}
\def\dt{\diff t}

\def\half{\frac{1}{2}}

\begin{document}
\title{Studies of Spuriously Shifting Resonances in Time-dependent Density Functional Theory}

\author{Kai Luo}
\affiliation{Department of Physics and Astronomy, Hunter College and the Graduate Center of the City University of New York, 695 Park Avenue, New York, New York 10065, USA}
\affiliation{Department of Chemistry and Chemical Biology, Cornell University, 259 East Ave, Ithaca, NY 14850, USA}

\author{Johanna I. Fuks}
\affiliation{Department of Physics and Astronomy, Hunter College and the Graduate Center of the City University of New York, 695 Park Avenue, New York, New York 10065, USA}

\author{Neepa T. Maitra}
\affiliation{Department of Physics and Astronomy, Hunter College and the Graduate Center of the City University of New York, 695 Park Avenue, New York, New York 10065, USA}

\date{\today}

\begin{abstract}
Adiabatic approximations in time-dependent density functional theory (TDDFT) will in general yield unphysical time-dependent shifts in the resonance positions of a system driven far from its ground-state. This spurious time-dependence is explained in [J. I. Fuks, K. Luo, E. D. Sandoval and N. T. Maitra, Phys. Rev. Lett. {\bf 114}, 183002 (2015)]   in terms of the violation of an exact condition by the non-equilibrium exchange-correlation kernel of TDDFT.  Here we give details on the derivation and discuss reformulations of the exact condition that apply in special cases.  In its most general form, the condition states that when a system is left in an arbitrary state,  the TDDFT resonance position for a given transition in the absence of time-dependent external fields and ionic motion,  is independent of the state.  Special cases include the invariance of TDDFT resonances computed with respect to any reference interacting stationary state of a fixed potential, and with respect to any choice of appropriate 
stationary Kohn-Sham reference state.  We then present several case studies, including one that utilizes the adiabatically-exact approximation, that illustrate the conditions and the impact of their violation on the accuracy of the ensuing dynamics. In particular, charge-transfer across a long-range molecule is hampered, and we show how adjusting the frequency of a driving field to match the time-dependent shift in the charge-transfer resonance frequency, results in a larger charge transfer over time.
\end{abstract}

\maketitle 

\section{Introduction}
\label{sec:introduction}
The art of making approximations in the {\it ab initio} quantum theory of many-body systems enables us to investigate
realistic systems of various sizes, ranging from atoms to biological molecules with affordable
computational resources.  Accurate but efficient approximations are crucial to reproduce the experimental results, improve our 
understanding of the mechanisms in play and make both qualitative and quantitative predictions.
Rapid progress in experimental spectroscopy has begun to unveil
dynamics of the electronic degrees of freedom on the timescale of tens of attoseconds, 
which provides a touchstone for the assessment of {\it ab initio} electronic
structure theories\cite{KI09,Goulielmakis10,Krauszetal15}.

Based on Runge-Gross theorem, time-dependent density functional theory(TDDFT) is an in-principle exact and
efficient theory\cite{RG84,TDDFTBook,CarstenBook}, that, with functional approximations inherited  from ground state density
functional theory stands out
prominently among all other methods. The interacting one-body density
$n(\br,t)$ is obtained from the evolution of the Kohn-Sham (KS)
system, a system of fictitious non-interacting particles.  The KS
particles evolve under a one-body potential $v\s(\br, t)$, following
$i \partial_t \varphi_k(\br,t)= (-\frac{\nabla^2}{2} +
v_s(\br,t))\varphi_k(\br,t)$, such that the interacting density is
reproduced from $n(\br,t)=\sum_k^{\mathrm{occ}}|\varphi_k(\br,t)|^2$.  The
theory is in principle exact but the KS potential $v_s(\br,t)$
contains an unknown contribution called exchange-correlation (xc)
potential $v\xc(\br, t)$, which in practice needs to be approximated.
The latter is a functional of the density at all points in space and
at all previous times as well as the initial interacting state $\Psi_0,$ and
initial KS state $\Phi_0$: $v\xc[n;\Psi_0, \Phi_0](t)$.

Most TDDFT calculations however are adiabatic, i.e. the instantaneous
density is plugged into a ground state functional, $v\xc^{\rm
  adia}[n;\Psi_0, \Phi_0](t) = v\xc^{\rm g.s.}[n(t)]$, neglecting all
dependence on the history of the density and on the initial states
$\Psi_0$ and $\Phi_0$.  Adiabatic linear response TDDFT is extensively
and successfully used to compute excitation energies of molecules and
solids\cite{TDDFTBook}
, but TDDFT is not limited to linear response:
the theorems state that the density response to any order in an
external perturbation can be reproduced.  It is a promising candidate
for modeling non-equilibrium electron dynamics since it can capture
the correlation effects with a relatively cheap computational cost.
Although the performance of available
functionals for non-equilibrium dynamics has been much less explored, promising results 
have been reported \cite{SYKIOB12,Retal13, Fetal13, WDRS15, WNSPWLTYB15}. 
On the other hand, work on small systems where numerically-exact or high-level wavefunction methods are applicable, has shown
that the approximate TDDFT functionals can yield significant errors in
the simulated dynamics~\cite{EFRM12,LFSEM14,TGK08,RG12,RB09,FERM13,FHTR11,RN11,HTPI14,RN12,RN12c,PHI15,FCG15, FLSM15}.
There is an urgent need to better understand the origin of errors in TDDFT
approximations in this realm, especially since many topical applications such as for example 
charge-transfer dynamics and excitonic coherences in light-harvesting systems \cite{Retal13,SFCG11,ScholesJCPL14} or 
attosecond control of electrons in real time \cite{Cavalieri07,KI09} to name a few, involve tracking the 
system as it evolves far from its ground state.
When simulating these experiments with TDDFT,
one particular problem that has come to light in a number of recent studies, is
that approximate TDDFT can present  
spurious time-dependence in the resonance positions in non-perturbative dynamics and discrepancies between resonant frequencies
computed from different reference states \cite{GBCWR13,HTPI14,FLSM15,PHI15, OMKPRV15, FCG15,RN12, RN12b, RN12c}.
Such an artifact can lead to inaccurate dynamics and can muddle the analysis of electron-ion interactions, coherent processes and 
quantum interferences, among others \cite{Retal13, Fetal13,SWB04, SWB10, SBW11, MLR14}. 
The problem is not unique to TDDFT: it is inherent to any method that utilizes 
approximate potentials that depend on the time-evolving orbitals, 
such as time-dependent
Hartree-Fock(TDHF) for instance \cite{PN08}.

Given that TDDFT is formally exact, there is hope that approximate functionals can be designed that minimize the spurious time-dependence of 
the resonances and thereby improve the accuracy of the predicted dynamics. To this end, the exact condition derived in Ref.~\cite{FLSM15} should be useful, and here we give details elaborating on the condition and its implications in various cases.

Any system can be perturbed continuously and left in a non-equilibrium state.
If we turn off the perturbation, it starts to evolve freely, the state being a superposition 
of the eigenstates of the static 
potential. In this work, we define resonances, or response frequencies, 
of the system via the poles of the density-response of the field-free system;
these appear as peaks in its absorption spectra.
From the principles of quantum mechanics, in the absence of ionic motion the  resonances are
independent of the instantaneous state of the system once any external field is turned off, and independent of the time the field is turned off.
The linear response of the system may show dynamics in the oscillator strengths, and new peaks may appear or disappear, but their  positions remain constant.
An exact TDDFT simulation 
reproduces the physical resonances at all times, but approximate xc functionals in general display spuriously time-dependent resonances, also referred to as ``peak-shifting''. 
Ref.~\cite{FLSM15} rationalized the spuriously time-dependent
resonances in terms of the violation of an exact condition on the xc functional.
Here we elaborate on the derivation and provide a detailed analysis of examples that illustrate this effect.

The paper is organized as follows. In Section~\ref{sec:non_eq_response} we review the generalized 
non-equilibrium linear response formalism for TDDFT, introduced in
Refs.~\cite{FLSM15, PS15},
pointing out differences from the standard ground state linear response.
We discuss the difficulty of defining a pole structure for the non-equilibrium KS response function around an arbitrary state.  
In Section~\ref{sec:ex_cond} we discuss in detail the exact condition presented in Ref.~\cite{FLSM15} and reformulate it in several ways, including the special case of response around a stationary state.
In Section~\ref{sec:realspace} we show explicit time-dependent electronic spectra computed using approximate functionals for charge transfer dynamics. 
In order to illustrate the non-adiabatic nature of this spurious ``peak shifting",
in Section~\ref{sec:realspace} we analyze the electronic spectra for adiabatic dynamics using the exact ground state functional.
For this purpose we utilize a lattice model system,
for which the exact ground state xc functional can be computed and then used to propagate the KS system. 
For the same system, we present a proof of concept showing the impact of the violation of the exact condition on the simulated dynamics. 
We conclude and outlook the future work in Section~\ref{sec:conclusions}.

\section{Non-equilibrium Response Theory}
\label{sec:non_eq_response}
\subsection{General Formalism}
\label{sub:General formalism}

The non-equilibrium response function plays a crucial role in
analyzing non-equilibrium dynamics, as has been recently highlighted
in theoretical formulations of time-resolved photoabsorption
spectroscopy~\cite{PS15,PSMS15}. In fact the language of pump-probe
dynamics is very useful for establishing the exact condition, although
the relevance of the condition is not at all restricted to such a setup.

\begin{figure}
\centering
\includegraphics[width=0.5\textwidth]{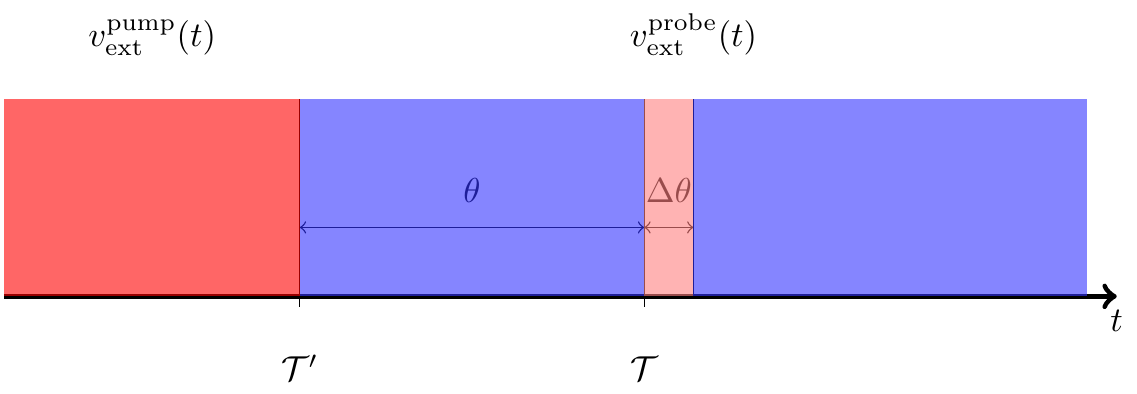}
\caption{The cartoon shows the time series of the applied field in a typical
    pump-probe experiment. Note that $v\ext^{(0)}$ is not shown since it is always on. At
    time $\mT'$, the pump is turned off and after a delay $\theta$, a weak probe laser
    is turned on for duration of $\Delta \theta$. In pump-probe spectroscopy, the experiment is repeated with different
    delays using the same probe. At $\mT + \Delta \theta$, the probe field has died off and only the static
    nuclear field is present. The times shaded in blue represents times when
    only the static potential $v\ext^{(0)}$ is present. 
    %When the pump dies off, the field becomes static. 
    %When the probe is weak enough, the external potential can be regarded as static.
}
\label{fig:pump-probe}
\end{figure}

Consider then a pump pulse that drives the system out of its ground state. At
time $\mT'$ the pump pulse is turned off and is followed, after some delay
$\theta$, by a weak probe pulse for duration $\Delta \theta$, see Figure~\ref{fig:pump-probe}. 

%\begin{figure}
%\begin{center}
%    \begin{tikzpicture} [scale=0.7, transform shape]
%        %\draw [->, ultra thick, opacity=1.0] (-5.0, 0) -- (0.0, 0) node[label={[shift={(0.0, 0.7)}]$\theta$}] {} -- (6.2, 0) node[anchor=north] {$t$} ;
%        \draw [->, ultra thick, opacity=1.0] (-5.0, 0)  -- (6.2, 0) node[anchor=north] {$t$} ;

%        \draw [] node[label={[shift={(-3.5, 2.2)}]$v_{\rm ext}^{\rm pump}(t)$}]  {};
%        \draw [] node[label={[shift={( 2.0, 2.2)}]$v_{\rm ext}^{\rm probe}(t)$}]  {};
 %       \draw [] node[label={[shift={( 3.5, 2.1)}]$v_{\rm ext}^{(0)}$}]  {};
 %       \draw [<->] (-2.0, 0.7) -- (1.5, 0.7) node[label={[shift={(-1.75, 0.0)}]$\theta$}]  {};
 %       \draw [<->] (1.5, 0.7) -- (2.0, 0.7) node[label={[shift={(-0.25,
 %       0.0)}]$\Delta\theta$}]  {};

 %       \draw [shift={(-2, 0)}] (0pt, 0pt) -- (0pt, 2) ;
 %       \draw [shift={(1.5, 0)} ] (0pt, 0pt) -- (0pt, 2) ;
 %       \draw [shift={(2, 0)} ] (0pt, 0pt) -- (0pt, 2) ;

 %       \draw [shift={(-2, 0)} ] (0pt, 0pt) -- (0pt, -3pt) node[label={[shift={(0.0, -1.0)}]$\mathcal{T}'$}]  {};
 %       \draw [shift={(2, 0)} ]  (0pt, 0pt) -- (0pt, -3pt) node[label={[shift={(0.0, -1.0)}]$\mathcal{T}$}]  {};

 %       \path[fill=red, opacity=0.6] (-5, 0pt) -- (-2, 0pt) -- (-2, 2.0) -- (-5, 2.0);

 %       \path[fill=blue!80, opacity=0.6] (-2, 0pt) --  (1.5,  0pt) -- (1.5, 2.0) -- (-2, 2.0);

 %       \path[fill=red!50, opacity=0.6] (1.5, 0pt) --  (2,  0pt) -- (2, 2.0) -- (1.5, 2.0);

 %       \path[fill=blue!80, opacity=0.6] (2, 0pt) --  (6,  0pt) -- (6, 2.0) -- (2, 2.0);

 %   \end{tikzpicture}
%\end{center}

The response of the excited system is monitored by repeating the experiment for different delay $\theta$. 
When the electronic response is measured, coupling of the electronic dynamics to ionic motion manifests 
itself as changes in the position of the spectral peaks with respect to $\mT'$ and $\theta$. 
But when ions are clamped the peak positions should not move; they should be independent of both $\theta$ and $\mT'$.
For $t>\mT'$ and considering the ions clamped within the timescale of interest,
the unperturbed Hamiltonian becomes static $\hat{H}^{(0)} = \hat{T} + \hat{W} + \hat{v}\ext^{(0)}$, with $\hat{H}^{(0)} \Psi_n= E_n \Psi_n$.
Throughout the paper, the superscript ${(0)}$ indicates a quantity in the absence of time-dependent external fields.
Here $\hat{T}$ and $\hat{W}$ are the kinetic energy and the electron-electron interaction energy operators respectively.
The system is left in a superposition state which, at times $t$ greater than
$\mT = \mT' +\theta$, in the {\it absence} of any probe pulse can be written,
\ben
 \Psi^{(0)}(t\ge\mT) = \sum_n c_n(\mT)\Psi_n e^{-iE_n(t-\mT)},
\label{eq:Psi_mT}
\een
with $c_n(\mathcal{T})= c_n(\mT') e^{-i E_n \theta}$ and the $(0)$ superscript denotes a field-free evolution.
We denote the time-dependent density of this state as $n^{(0)}_\mT(\br,t)$, defined for times $t\ge\mT$,
\ben
n_\mT^{(0)}(\br,t)= N \sum_{\sigma_1,...\sigma_N} \int
\dr_2...\dr_N |\Psi^{(0)}(\br\sigma_1,\br_2\sigma_2,...\br_N\sigma_N, t)|^2\,,
\label{eq:n}
\een
where the indices $\sigma$'s represent the spin.
The non-equilibrium response function, $\tilde{\chi}$(we distinguish it from the ground state response 
function $\chi$ by a tilde), describing the density response $\d n(\br,t)$ to a perturbation $\delta v\ext(\br',t')$ (probe) applied at time $t'<t$ reads, 
\ben
\tilde\chi[n^{(0)}_\mT; \Psi(\mT)](\br,\br',t,t') = \left.\frac{\d n(\br,t)}{\d v\ext(\br',t')}\right\vert_{n^{(0)}_\mT,\Psi(\mT)}\,.
\label{eq:chidef}
\een
In principle, $\tilde{\chi}$ depends on the unperturbed density $n^{(0)}(t)$ at times between $\mT$ and $t$, and on the state at time $\mT$, as follows from the  Runge-Gross theorem~\cite{RG84}.
Widely discussed in the literature is the ground state response function, 
which is a particular case of Eq.~(\ref{eq:chidef}) when the initial state is the ground state,
$\Psi(\mT)=\Psi_0$, and the unperturbed density becomes the ground state density  $n^{(0)}_\mT(\br,0)=n_0(\br)$.
The ground state response function $\chi[n_0]$ depends only on the time interval $\tau=t-t'$,
\ben
\chi[n_0](\br, \br',t,t')  = \left.\frac{\d n(\br,t)}{\d v\ext(\br',t')}\right\vert_{n_0}\equiv \chi_0(\br,\br', \tau)\;
\label{eq:chigs}
\een
due to the time-translation invariance of the ground state.
Fourier transforming with respect to $\tau$  yields the zero temperature Lehmann
representation, see e.g. Ref.~\cite{GVbook},
\ben
\begin{split}
    \chi_0(\br, \br', \w) & = \int\!\!  \diff\tau \chi_0(\br, \br', \tau)  e^{-i \w \tau} \\
     &  = \sum_{k} \left( \frac{f_{0k}(\br)f_{k0}(\br')}{\w - \w_{k0}+i0^+} - \frac{f_{0k}(\br')f_{k0}(\br)}{\w + \w_{k0}+i0^+}\right)\,.
\end{split}
 \label{eq:chi_0}
   \een
where $f_{jl}(\br) = \langle \Psi_j \vert \hat{n}(\br) \vert \Psi_l \rangle$ and $\w_{jl} = E_j -E_l$.
$\chi_0(\br, \br', \w)$ has poles at the frequencies 
corresponding to transitions from the ground state to other eigenstates $k$ of the system. 

Unlike the ground state response function the generalized non-equilibrium response function $\tilde\chi[n^{(0)}_\mT; \Psi(\mT)]
(\br,\br',t,t')$ defined in Eq.~(\ref{eq:chidef})
 in principle  depends both on $t$ and $t'$ independently.
Following derivations in standard linear response theory~\cite{GVbook} but now generalized to an arbitrary initial state:
\bea
&&  \tilde\chi[n^{(0)}_\mT; \Psi(\mT)](\br,\br',t,t')
\nonumber
\\
& =&  -i \theta(t-t') \braket{\Psi(\mT) |[\hn(\br,t), \hn(\br',t')] | \Psi(\mT)}. 
\label{eq:chi_tilde}    
\eea
In Eq.~(\ref{eq:chi_tilde}) the density operator is in the interaction picture: $\hn(\br,t) = e^{i\hH^{(0)} t} \hn(\br)
e^{-i\hH^{(0)}t}$. Given that the response function is defined as the response
of the density at time $t$ with respect to a perturbation at time $t'$
we define again $\tau = t-t'$, which we will Fourier transform with respect to,
in order to obtain a spectral representation.
Then, inserting the expansion for the arbitrary state Eq.~(\ref{eq:Psi_mT}) in Eq.~(\ref{eq:chi_tilde}), we have 
\begin{widetext}
\begin{eqnarray}
\tilde{\chi}_{t'}(\br, \br', \tau) = -i \theta(\tau) \sum_{m,n,k} P_{nm}(\mT) 
e^{i \w_{nm} t'}  \left[ e^{i\w_{nk}\tau} f_{nk}(\br) f_{km}(\br') -
    e^{i\w_{km}\tau } f_{nk}(\br') f_{km}(\br) \right]
\end{eqnarray}
with 
$P_{jl}(\mT) = c_j^*(\mT)c_l(\mT)$, where we choose to parameterize the response function via $t'$.
Instead, one could parameterize it via $T =(t+t')/2$, in which case we have,
\begin{eqnarray}
\tilde{\chi}_{T}(\br, \br', \tau) 
=-i\theta(\tau)\sum_{n,m} P_{nm}(\mT) e^{i\w_{nm}  T }\sum_k
\left[e^{i \frac{\w_{nk}+\w_{mk}}{2}\tau } f_{nk}(\br)f_{km}(\br') 
    -e^{-i \frac{\w_{nk}+\w_{mk}}{2}\tau } f_{nk}(\br')f_{km}(\br) 
\right]\,.
\end{eqnarray}
Performing the Fourier transform yields, for the two choices of parametrization,
\begin{eqnarray}
\tilde{\chi}_{t'}(\br,\br', \w) = \chi^{\mathrm{diag}}(\br,\br',\w) +
    \left( \sum_{n\neq m} P_{nm}(\mT) e^{i\w_{nm} t'} \sum_k\left[\frac{f_{nk}(\br) f_{km}(\br')}{\w -\w_{kn}
            +i0^+} - \frac{f_{nk}(\br') f_{km}(\br) }{\w + \w_{km} +i0^+}
    \right]\right)\,,
\label{eq:chi1(om,t')}
\end{eqnarray}
and
\begin{eqnarray}
\tilde\chi_{T}(\br,\br',\w) = \chi^{\mathrm{diag}}(\br,\br',\w) + 
 \left( \sum_{n\neq m} P_{nm}(\mT)e^{i\w_{nm} T} \sum_k\left[
\frac{ f_{nk}(\br)f_{km}(\br')}{\w - \frac{\w_{kn} + \w_{km}}{2} +i0^+} - 
 \frac{ f_{nk}(\br')f_{km}(\br)}{\w + \frac{\w_{kn} + \w_{km}}{2} +i0^+}
 \right]\right) \,,
\label{eq:chi2(om,T)}
\end{eqnarray}
where the diagonal term corresponding to $m=n$, an incoherent sum over populations of initially occupied states, has been isolated,
\ben
\chi^{\mathrm{diag}}(\br,\br',\w) = \sum_n P_{nn}(\mT) \sum_k  \left(\frac{f_{nk}(\br) f_{kn}(\br')}{\w -\w_{kn} +i0^+} -
    \frac{f_{nk}(\br') f_{kn}(\br) }{\w + \w_{kn} +i0^+} \right)
\label{eq:diag}
\een
Note that the second terms in the large curved parenthesis on the right of Eqs.~(\ref{eq:chi1(om,t')}), (\ref{eq:chi2(om,T)}), and~(\ref{eq:diag}), are simply complex conjugates of the first terms when evaluated at $-\omega$; in the following we will replace such terms by the expression $+ c.c.(\omega \to -\omega)$.  It may appear from the different pole structure of Eqs.~(\ref{eq:chi1(om,t')}) and~(\ref{eq:chi2(om,T)}), that $\tilde\chi_{t'}$ and $\tilde\chi_{T}$ yield different density-responses, but in fact this is not the case. Since they originate from the same  $\tilde{\chi} (\br,\br',
t,t')$ they must yield the same density-response $\d n(\br, \w)$, and we now show this explicitly;
in particular the poles at half-sum frequencies in $\tilde\chi_{T}$ vanish once $\d n(\br,\w)$ is computed.
The density-response in the frequency domain $\d n(\br,\w) = \int\!\!\dt\; \d
n(\br,t)e^{i\w t}$ is computed via 
\ben
 \d n(\br,t) = \int\!\! \dr' \dt' \; \tilde\chi(\br, \br', t, t') \d v(\br', t')\,.
\label{eq:den_resp}
\een
Taking the inverse Fourier transforms, $\tilde\chi_{t'}(\br, \br', \tau) = \frac{1}{2\pi}  \int\! \diff w_1\, e^{-i\w_1 \tau} \tilde \chi_{t'}(\br, \br', \w_1)$ 
and $\d v(\br',t') =\frac{1}{2\pi} \int\! \diff \w_2\, e^{-i\w_2 t'} \d v(\br', \w_2)$,
we have
\ben
\d n(\br,\w) = \int\!\!\dr'\dt \dt' \frac{\diff\w_1}{2\pi} \frac{\diff \w_2}{2\pi} e^{i \w t}  e^{-i\w_1 \tau}  
e^{-i\w_2 t'} \tilde\chi_{t'}(\br, \br', \w_1) \d v(\br', \w_2)\,.
\label{eq:dens_resp_freq}
\een
Inserting Eq.~(\ref{eq:chi1(om,t')}) into Eq.~(\ref{eq:dens_resp_freq}) yields 
two contributions:
\ben
\delta n(\br,\w) = \d n^{\mathrm{diag}}(\br,\w) + 
\int\!\! \dr'  \left[\sum_{n\neq m} P_{nm}(\mT) \sum_k\frac{f_{nk}(\br) f_{km}(\br')}{\w -\w_{kn}
        +i0^+} + c.c. (\omega \to -\omega)
\right] \d v(\br', \w + \w_{nm})\,,
\label{eq:dn_omega}
\een
where
\ben
\d n^{\mathrm{diag}} (\br,\w) = \int \!\dr'\, \sum_n P_{nn}(\mT) \sum_k  
\left[\frac{f_{nk}(\br) f_{kn}(\br')}{\w -\w_{kn} +i0^+} + c.c. (\omega \to -\omega)
\right] \d v(\br',\w)\,.
\label{eq:dn_omega_D}
\een
\end{widetext}
The first term in Eq.~(\ref{eq:dn_omega}) arises from the diagonal term, where the 
poles are at the true excitations: the energy differences between an occupied state $n$ and an unoccupied state $k$. The second term arises from the 
second term  of Eq.~(\ref{eq:chi1(om,t')}); the identities $\int\! \dt\, e^{i\w t}
e^{-i\w_1 t} = 2\pi\delta(\w-\w_1)$ and  $\int \dt' e^{i\w_1t'} e^{-i\w_2 t'} e^{i\w_{nm} t'}= 2\pi\delta(\w_1-\w_2+\w_{nm})$ 
enabled us to readily perform the integrals.
If instead of $\chi_{t'}$, we insert  $\chi_{T}$ into Eq.~(\ref{eq:dens_resp_freq}),
we find exactly the same expression as Eq.~(\ref{eq:dn_omega}) after recognizing that
the $\dt$ and $\dt'$ integrals in this case yield $\d(\w- \w_1 + \w_{nm}/2)$ and
$\d(\w_2 - \w_1 - \w_{nm}/2)$ respectively. 

The diagonal term has the usual structure showing resonant peaks where
the perturbing potential has components at a frequency equal to an
energy-difference between occupied and unoccupied states; the only
difference with the ground-state response is that states other than
the ground-state are occupied, as reflected in the $P_{nn}(\mT)$
prefactor. The off-diagonal term also has response peaks at the
occupied--unoccupied energy differences: the peaks are at the zeroes
of the denominator, at frequencies $\omega_{kn}$. Considering that
this is multiplied by the $(\omega +\omega_{nm})$-frequency-component
of the applied potential $\delta v$, then we see that this term is
resonant when the applied potential resonates at $\omega_{kn} +
\omega_{nm} = \omega_{km}$.

\subsection{Generalized Non-Equilibrium Kohn-Sham Response}
\label{sec:KS_response}
Generally the wavefunctions that are summed over in response function of an interacting system are
inaccessible for a realistic system, so one turns to alternative methods to calculate it, such as TDDFT.
Unlike the interacting system, which, after the field is turned off evolves
under the {\it static} potential $v\ext^{(0)}(\br)$, 
the KS system evolves in the potential
\ben
\begin{split}
v\s[n_\mT^{(0)},\Phi(\mT)](\br,t) & = v\ext^{(0)}(\br) +  v\Hxc[n_\mT^{(0)}; \Psi(\mT),\Phi(\mT)](\br,t)
\\ & \equiv v\s^{(0)}(\br,t),
\end{split}
\label{eq:vs}
\een
which typically continues to evolve in time for $t \geq \mT$ even in the absence of time-dependent external fields.
The time-dependence of the KS potential $v\s^{(0)}[n_\mT^{(0)}, \Phi(\mT)](\br,t)$ is due 
to the potential itself being a functional of the non-stationary density
$n_\mT^{(0)}(\br, t)$ and the KS initial state $\Phi(\mT)$.

Similar as for the ground state response $\chi_0$, TDDFT can be used to find the
interacting non-equilibrium response $\tilde{\chi}$.
Since the physical and KS system must yield the same density-response, we can
derive  a Dyson-like equation linking the generalized non-equilibrium interacting response function
$\tilde\chi$ Eq.~(\ref{eq:chidef})
with a generalized non-equilibrium KS response function $\tilde\chi\s$, 
\ben
\tilde\chi^{-1}[n^{(0)}_{\mT}; \Psi(\mT)]=
\tilde\chi\s^{-1}[n^{(0)}_{\mT}; \Phi(\mT)]  - \tilde{f}\Hxc[n^{(0)}_\mT; \Psi(\mT), \Phi(\mT)]\,
\label{eq:DyEq}
\een 
via a generalized Hartree-xc kernel $\tilde f\Hxc = 1/\vert\br - \br'\vert + \tilde f\xc$, 
with,
\ben
\tilde{f}\xc[n^{(0)}_\mT;\Psi(\mT),\Phi(\mT)](\br,\br',t,t') = 
\frac{\delta v\xc(\br,t)}{\delta n(\br',t')} \Big\vert_{n^{(0)}_\mT,\Psi(\mT),\Phi(\mT)}.
\label{eq:fxc}
\een
Note that the dependence on the  unperturbed density $n^{(0)}$ at times between $\mT$ and $t$, and
the interacting and KS initial states $\Psi(\mT)$, $\Phi(\mT)$ in
$\tilde\chi(\br,\br',t,t')[n^{(0)}_{\mT}; \Psi(\mT)]$ and $\tilde\chi\s(\br,
\br', t,t')[n^{(0)}_{\mT}; \Phi(\mT)]$ follows from the Runge-Gross and van
Leeuwen proofs~\cite{RG84,L99}.
As a consequence of the time-dependence of $v\s^{(0)}$ 
the bare KS eigenvalues and eigenvalue-differences also become time-dependent
\footnote{The time-dependent KS eigenvalues are referred as the eigenvalues when
one solves a static KS equation with the instaneous xc potential.}.
%and do not coincide with the KS response frequencies.
%The later are defined as the poles of the $\tau$-Fourier transform of the KS response function:
The KS response function in Eq.~(\ref{eq:DyEq}) is defined as
\ben
\tilde\chi\s[n^{(0)}_\mT,\Phi(\mT)](\br, \br', t,t') = \left.\frac{\delta
        n(\br,t)}{\delta v\s(\br',t')}\right \vert_{n^{(0)}_\mT,\Phi(\mT)}
        \equiv \tilde{\chi}\s(\br, \br',t, t')\,.
\label{eq:chis}
\een
%Similar to the definition of the interacting response function, 
Unlike $\tilde \chi$,  $\tilde \chi\s$ in general does not have a simple spectral (Lehmann) representation.
%like Eq.~\ref{eq:den_resp}. 
The reason is the interaction picture for the KS system involves a
time-dependent Hamiltonian, $H\s^{(0)}(t)= -\nabla^2/2 +
\hat{v}\s[n_\mT^{(0)},\Phi(\mT)](t)$, therefore the non-equilibrium KS
response~\cite{UGG95,GVbook},
\bea
   && \tilde{\chi}\s(\br,\br',t, t') =  -i \theta(t-t') \braket{\Phi(\mT) |[\hn(\br,t), \hn(\br',t')] | \Phi(\mT)} 
  \nonumber \\
  &&= -i \theta(t-t') \sum_{l,k} \varphi_{k}^{*}(\br,t)
\varphi_{l}^{*}(\br',t') \varphi_{k}^{*}(\br',t')\varphi_{l}(\br,t) + c.c.\,.
\label{eq:chis_2}    
\eea
involves time-ordered operators 
$\hat{n}(\br,t)  = \hat{T}e^{i\int_0^t d\tau   \hH\s^{(0)}(\tau)}\hat{n}(\br)\hat{T}e^{-i\int_0^t d\tau  
    \hH\s^{(0)}(\tau)}$, with $\hat{T}$ being the time-ordering operator. 
A simple interpretation of the Fourier transform of $\tilde{\chi}\s(\br,\br',t, t')$ with respect to $\tau$, $\tilde\chi_{{\sss{S}},t'}(\br,\br',\w)$  or
$\tilde\chi_{{\sss{S}},T}(\br,\br',\w)$, as in the SEc. \ref{sec:non_eq_response},
in terms of eigenvalue differences of some static KS Hamiltonian is generally not possible. 
%The time-dependence of the instantaneous KS eigenvalues prevent us from pulling down terms like $\w \pm
%\w_{ij}$ within the Fourier transform. 
%Actually, we couldn't find a proof that $\tilde\chi_{{\sss{S}},[t',T]}$ has poles in the general case, but only for particular cases, as will be discussed in next section. 
Despite the fact that a pole structure for $\tilde{\chi}\s$ may not be simple to define in general (see
particular cases for which it can be done in Sec. \ref{sec:realspace}), when $\tilde{\chi}$ is constructed from
$\tilde{\chi}\s$ and $\tilde{f}\Hxc$ via Eq.~(\ref{eq:DyEq}), the interacting
spectral representation is retrieved.

\section{Exact Conditions}
\label{sec:ex_cond}
As discussed in Sec. \ref{sec:KS_response} the KS potential for non-stationary dynamics Eq.~(\ref{eq:vs}) is
generally time-dependent even after any external field is turned off. This is the case both for the exact KS potential and also for adiabatic approximations, because the density $n^{(0)}_\mT$ is time-dependent. 
The instantaneous eigenvalue-differences of the KS Hamiltonian are time-dependent because $v\Hxc[n_\mT^{(0)}; \Psi(\mT),\Phi(\mT)]$ changes as the density evolves.
For ground state linear response, $f\xc$ must shift the KS response frequencies to the physical ones
and create missing multi-electron excitations. For the present non-equilibrium
response, the generalized $\tilde{f}\xc$ must additionally cancel 
spurious $\mT$-dependence in Eq.~(\ref{eq:DyEq}) to ensure
$\mT$-{\it independent} TDDFT resonances.  Cancellation of spurious
$\mT$-dependence is an exact condition on the xc functional
\cite{FLSM15}.  We give the general form of this
condition in (i) below, and in (ii) and (iii) we discuss implications for a few
special cases.

(i) {\it Condition 1: Invariance with respect to $\mT$.}

Consider the TDDFT prediction for the transition frequency $\w_i$ between two given interacting states.
This $\w_i$ is a pole of 
\ben
\left(\tilde\chi\s^{-1}[n^{(0)}_\mT,\Phi(\mT)] - \tilde f\Hxc[n^{(0)}_\mT;
    \Psi(\mT), \Phi(\mT)]\right)^{-1}\;.
\een 
Then $\w_i$ should be invariant with respect to $\mathcal{T}$:
\ben  
\frac{\diff \w_i}{\diff \mT} = 0\;.
\label{eq:ex_cond}
\een 
We shall call this {\it Condition 1} in the following. 

We note that as a system evolves under an external field, more states
can become populated and so when the field is then turned off and the
system's linear response probed, new frequencies may appear corresponding to
transitions from states
populated at the later time that were not populated at earlier
times. Likewise, some frequencies appearing in the response at earlier
times may disappear. The exact condition addresses those response frequencies present at both earlier and later times: the positions of these must be invariant. 
(See also consistency condition discussed in  {\it ii}).
Although expressed above in the context of pump-probe
spectroscopy, the exact condition clearly applies to any
non-equilibrium dynamics, where $\w_i$ are the field-free
resonance positions of the system at some time $\mT$. It can be viewed
in another way: let $\Psi_\mT$ be {\it any} arbitrary interacting
state, not necessarily a stationary state, of a system in a static
potential $v\ext^{(0)}$, and $n^{(0)}_\mT(t)$ be its field-free
time-dependent density as it evolves in $v\ext^{(0)}$. The subscript $\mT$
is no longer a time label but instead labels a particular arbitrary
state $\Psi_{\mT}$. Then the response of this arbitrary state has poles at
frequencies $\w_i$, which for a given transition satisfy Eq.~(\ref{eq:ex_cond}): their positions
are independent of the choice of this arbitrary state, i.e of $\mT$.

(ii) {\it Condition 2: Invariance with respect to any stationary
        interacting state of a given static potential}

Consider now the special case when the interacting system is in
a stationary excited state, $\Psi(\mT)= \Psi_k$ of the static
potential that it lives in, $v\ext^{(0)}$.  The density is then
stationary, $n^{(0)}_\mT(t)=n_k$. The initial state chosen for the KS
calculation must have the same density and its first time-derivative
as the initial interacting state~\cite{L99,MB01}, but need not itself
be a stationary state of any potential, since it is possible for
time-dependence of different KS orbitals to cancel each other
out. However, let us choose a KS initial state $\Phi_l$ with density
$n_k$ that is a stationary excited state of some static one-body
potential $v\s^{(0)}$. 
(Note that $v\s^{(0)}$ is {\it not} the
ground-state KS potential corresponding to the external potential
$v\ext^{(0)}$; they do not have the same ground-state density, but
rather the excited state $\Phi_l$ of the former has the same density
$n_k$ as the excited state $\Psi_k$ of the latter). 
Then, because the
unperturbed KS system is static, we can find a Lehmann representation
for the KS density-response function, and derive a matrix formulation
of the Dyson equation Eq.~(\ref{eq:DyEq}) in frequency-domain. The poles
of the frequency-domain Dyson equation are eigenvalues of the matrix equation
%, similar to the procedure in Ref.~
 \cite{GPG00}.

Now let $\{\Psi_1,...\Psi_k...\}$ be a set of eigenstates
of the same static potential
$v\ext^{(0)}$, and imagine finding an appropriate KS stationary state $\Phi_{k}$
for each one, that is an excited state of some KS potential $v\s^{(0)}[n_k;\Phi_k]$, reproducing the
density, $n_k$, of the $k$th true excited state $\Psi_k$. 
Note that, although we use here the same index $k$ to label the corresponding 
KS wavefunction instead of $l$ as in the previous paragraph, to avoid proliferation of subscripts, it need not be the $k$th KS excitation of 
the potential in which it is an eigenstate, $v\s[n_k; \Phi_k]$ (see also {\it   Condition 3} (iii) shortly).
Then the exact condition is that the TDDFT response frequency predicted for a given transition between two states $\Psi_{k}$ and $\Psi_{k'}$ of a given potential $v\ext^{(0)}$ must be
independent of which of these two states is chosen as the reference, i.e. the TDDFT response frequency should be identical whether the response is calculated around $\Psi_{k}$ or around $\Psi_{k'}$. 
We call this {\it Condition 2} in the following.  Other excitation frequencies
of the system must be {\it consistent}.  For example, consider computing the
response around (a) state $\Psi_2$ and (b) state $\Psi_3$. Then this condition
expresses that $\vert \omega_{32}^{a}\vert$ predicted in calculation (a) must
equal  $\vert \omega_{23}^{b}\vert$ predicted in calculation (b), but {\it also}
other frequencies have to be {\it consistent}, i.e. we must have $\omega_{k3}^{(b)} = \omega_{k2}^{(a)} - \omega_{32}^{(a)}$.  
Within a single-pole approximation (SPA) of this
generalized Dyson equation (justified for a KS transition well-separated from all other transitions~\cite{PGG96}), {\it Condition 2} simplifies to the condition
that,
\ben
\w^{\rm SPA} = \w_{{\sss{S}},q}^{k} + 2 \Re \int\!\! \dr \int\!\! \dr'
\bar{\Phi}_q^{k}(\br) \tilde{f}^k\Hxc(\br,\br',\w_{{\sss{S}},q}^{k}) \Phi_q^k(\br')
\label{eq:SPA}
\een
is the same whether $\Psi_{k}$ or $\Psi_{k'}$ is chosen as the reference $k$th interacting state.
The notation $\w_{{\sss{S}},q}^{k}$ means the $i\to a$ KS excitation out of the potential  $v\s^{(0)}[n_k;\Phi_k]$ and $q=(i,a)$ represents a double index.
%, computed around the interacting state $\Psi_{k}$. 
We have defined the transition density
\ben
\Phi_q^k(\br) = \bar{\varphi}_i^k(\br) \varphi_a^k(\br), 
\label{eq:transden}
\een
in which $\bar{\varphi}_i^k(\br)$ denotes the complex conjugate of $\varphi_i^k(\br)$.
Here $\varphi_i^{k}, \varphi_a^k$ are the initial occupied and unoccupied KS orbitals of 
potential $v\s^{(0)}[n_k;\Phi_k]$, and
$\w_{{\sss{S}},q}^{k}$ is their orbital energy difference. The
shorthand $\tilde{f}\Hxc^{k}$ represents the generalized kernel
$\tilde{f}\Hxc[\Psi_k,\Phi_k]$ (the field-free density-dependence is
redundant when beginning in a specified stationary state of the unperturbed
potential, since  that information is contained already in the initial states). 
 The expression
Eq.~(\ref{eq:SPA}) is given for the spin-saturated case; for
spin-polarized systems and non-degenerate KS poles, replace $\tilde
f\Hxc$ with $(1/\vert\br - \br'\vert +\tilde f\xc^{\sigma,\sigma'})$, where $\sigma$ and $\sigma'$ are spin indices.

{\it Condition 2} and Eq.~(\ref{eq:SPA}) were given in Ref.~\cite{FLSM15} but
discussed within the adiabatic approximation, where, having a static
density is enough to guarantee that the KS potential is static, and
$\Phi_k$ solves self-consistent field (SCF) equations for the static
potential $v\s^{(0)}[n_k; \Phi_k]$.

Like {\it Condition 1}, the
degree to which {\it Condition 2} is violated can be used to determine
the accuracy of the dynamics, especially relevant when the dynamics
involves just a few interacting excited states that get populated and
depopulated in time.  Strictly speaking, to use {\it Condition 2}, one
would need to know the exact density of the interacting excited
states, but in practise one can often find appropriate KS excited
states for a given functional approximation, whose densities are
assumed to approximate the interacting ones~\cite{EM12}. This was done in
Ref.~\cite{FLSM15} for a few examples (see also
section~\ref{sec:realspace}), and we will also utilize this
approximate {\it Condition 2} in the next section.

(iii) {\it Condition 3: Invariance with respect to the choice of a stationary
    non-interacting state $\Phi_k$.}

Continuing with response of stationary states, not only must the
TDDFT frequencies be independent of the interacting excited state of a
fixed potential, but also they must be independent of the choice of
the KS state. Let $\{\Phi_1,\Phi_2...\Phi_l...\}$, each with density
$n_k$ which is the density of a fixed interacting excited state $\Psi_k$ of a given external potential $v\ext^{(0)}$,
be possible KS stationary states of different one-body potentials $\{
v^{(0)}_{s,1},v^{(0)}_{s,2}...v^{(0)}_{s,l}\}$. Then {\it Condition 3} states that, for a given transition, the
frequency obtained via TDDFT response  must be the same for any of these $\Phi_l$'s.  
%We refer to this as {\it Condition 3}. 
Again, like
in (ii), one can express this condition directly in the
frequency-domain in a generalized matrix formulation which within the
single-pole approximation reduces to Eq.~(\ref{eq:SPA}) where here $k$
should be replaced by $l$, labelling the particular KS state around
which the TDDFT response is calculated.

\section{Spurious Time-dependent Spectra}
\label{sec:realspace}
We illustrate the significance of our exact conditions using two examples of
charge-transfer dynamics. Charge transfer(CT) is a crucial process in many
topical applications today, including photovoltaic design where the
system is initially photoexcited, and transport in nano-scale devices
where the system begins in its ground state.  For each of these
scenarios, we utilize model two-electron systems for which exact
results are available to compare with, and for which we can thoroughly
analyze how the spectral peaks shift as a function of time. We will
find both the {\it Conditions 1} and {\it Condition 2} 
are useful to understand the performance of the approximate
functionals. We also discuss a situation where {\it Condition 3} is violated.

\subsection{Charge Transfer from a Photoexcited State}
\label{sec:photoCT}
Our first example revisits an example of Ref.~\cite{FLSM15},
focusing here on the electronic spectra as the system evolves,
obtained via the unrestricted exact-exchange approximation (EXX) and self-interaction-corrected
local spin density approximation (SIC-LSD).  We consider two
interacting electrons in a one-dimensional double well described by,
\ben
v\ext(x) = -\frac{Z_L}{\sqrt{(x+R)^2 + 1}} -\frac{U_L}{\cosh^2(x+R)} - \frac{U_R}{\cosh^2(x-R)}
\label{ref:H}
\een
with parameters $Z_L = 2, U_L =2.9, U_R=1$ and $R = 3.5$
a.u. Electrons interact via soft-Coulomb $w(x',x)
=1/\sqrt{(x'-x)^2+1}$.  The spacing of the simulation box is $0.1$
a.u. and we use zero boundary conditions at $\pm 20.0$ a.u.
Atomic units are used throughout. The real-time propagation is done
using  real-space code {\sf octopus}~\cite{octopus,octopus2, octopus3} with a
time step of $0.005$ a.u..  

\begin{figure}[th]
    \centering
    \includegraphics[width=0.5\textwidth]{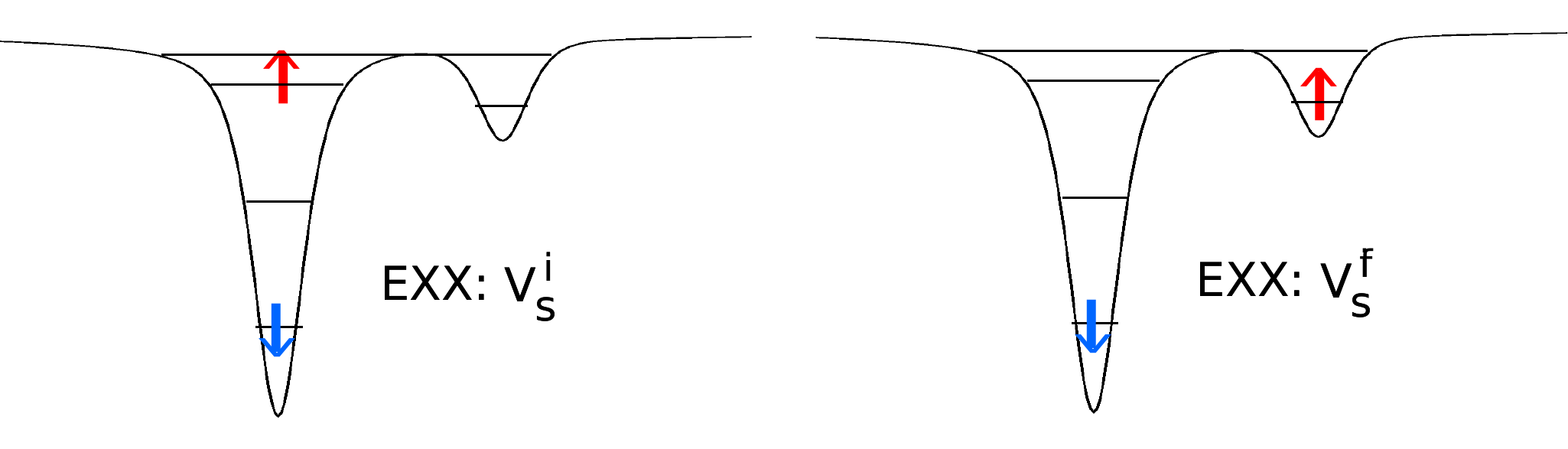}
    \caption{To the left the  EXX potential $v\s^{\uparrow,i}$ and the initial configuration for the photoexcited 
    KS state; to the right the final  EXX potential $v\s^{\uparrow,f}$ and the
    target CT state are plotted. Here the position of the arrow denotes where
    most of the density sits.
    The model is specified in Section \ref{sec:photoCT}.
    As discussed in the text for this photoexcited CT dynamics the EXX potential the transferring $\uparrow-$electron experiences 
    (in the absence of time-dependent external fields) is nearly constant.}
\label{fig:phCT}
\end{figure}

The interacting system is prepared in an excited eigenstate of the unperturbed Hamiltonian and then evolved in the
presence of a weak monochromatic laser resonant with the
photoexcited-to-CT transition frequency $\w_{CT}$. 
The latter corresponds to the energy difference between the initial photoexcited state, denoted by $i$,
and the final target CT state, denoted $f$, $\w_{CT}=E_f-E_i=0.289$~a.u..
The parameters are chosen such that Rabi oscillations between the photoexcited and the CT states are induced 
(although other states get lightly populated).  The exact interacting dynamics is compared
with the results from the various TDDFT approximations. For the
latter, the initial state is obtained by promoting a KS particle from the
ground state configuration to an unoccupied orbital (see the left-hand side of
Fig.~\ref{fig:phCT}), such that the
dominant configuration of the interacting initial state is mimicked. This KS
excited state is then relaxed via an SCF calculation to be the KS
eigenstate of the initial KS potential $v\s^{(0)}(\br,0)$, which
prevents any dynamics before the applied field is turned on.  The KS
system is propagated in the presence of a weak laser resonant with the
TDDFT CT resonance of the given approximate functional computed via
linear response around the initial KS photoexcited state. 
(As is common practise, spin-polarized dynamics is run from the initial singly-excited KS determinant). 
We note that
if {\it Condition 2} is violated, this value of
the CT resonance is {\it not} the value determined from the
usual linear response around the ground state, nor is it the value
determined from linear response around the target CT state. In fact, as
pointed out in Ref.~\cite{FLSM15}, the deviation of the resonance
predicted from linear response around the initial state, $\w_i$, to that from
target state, $\w_f$, is
strongly correlated with the performance of the functional approximation.
For example, for the two functionals we present here, for EXX $\w^i=\w^f =
0.287$a.u.  while for SIC-LSD  $\w^i =0.287$a.u. and $\w^f = 0.237$a.u.. 
Accordingly, we found in Ref.~\cite{FLSM15} that EXX demonstrated near-perfect charge-transfer, 
while SIC-LSD began promisingly but ultimately failed to transfer the charge. (Ref.~\cite{FLSM15} also considered LSD, whose discrepancy between the frequencies determined from the initial and target state responses was even greater, and its dynamics was even worse). 
For the same cases, we
demonstrate here explicitly the violation of {\it Condition 1}, namely the unphysical
time-dependence of the resonance positions, as a function of $\mT$.

At various times $\mT$ during the evolution we turn off the monochromatic laser and perform a linear density-response 
calculation to obtain the spectra at these times. The latter is done by applying a delta-kick~\cite{YNIB06} right after the laser is turned off, 
followed by a free evolution of some duration $T$ and then Fourier transforming the ensuing dipole difference between the kicked and un-kicked free propagations~\cite{RN12c,GBCWR13, OMKPRV15}
\ben
\Delta d(t) = d^{\mathrm{kicked}}(t)-d^{\mathrm{un-kicked}}(t)\,.
\label{eq:unkicked}
\een
We then plot the dipole spectrum $|\Delta d(\w)|$, denoted the ``absolute spectrum", 
\ben
|\Delta d(\w)| = \vert\int \!\dt\, e^{-i \w t} \Delta d(t)\vert\,,
\label{eq:delta_pow_spec}
\een
We evolved the field-free system  for $T=5000~a.u,$, resulting in a 
frequency-resolution of $2 \pi/ T \approx 0.001$ a.u.. 
We plot the absolute spectrum $|\Delta d(\w)|$ instead of the absorption spectrum $Im[d(w)]$ 
to simplify the analysis, since here we are only analyzing the position of the spectral peaks, not their oscillator strengths.

\begin{figure}[th]
    \centering
    \includegraphics[width=0.5\textwidth]{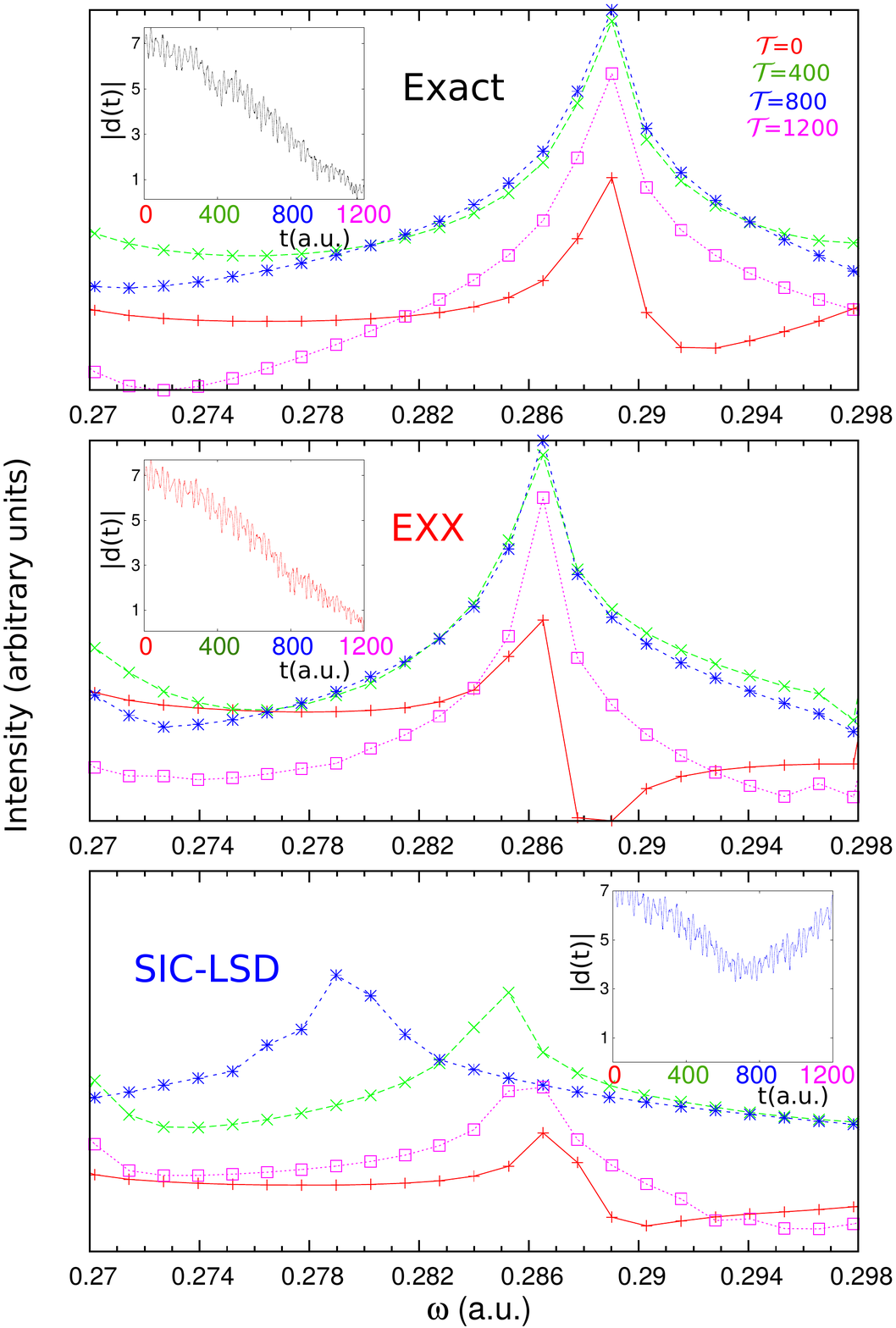}
    \caption{Logarithm of absolute spectrum $|\Delta d(\w)|$
        Eq.~(\ref{eq:delta_pow_spec})  showing CT frequency for different laser durations $\mT$. Inset: dipole moments $|d(t)|$ for the photoexcited CT dynamics studied in Ref.~\cite{FLSM15}. 
        Upper panel: Exact. Middle panel: EXX. Lower panel: SIC-LSD.}
\label{fig:ex_exx_sic_power}
\end{figure}

Fig.~\ref{fig:ex_exx_sic_power} shows the spectra for the exact,
EXX, and SIC-LSD cases, each driven at their respective initial response frequencies $\w^i$,
at different times $\mT=0, 400, 800, 1200$ a.u. of the photoexcited CT dynamics shown in
the inset. 
As expected, the exact CT resonance peak only changes strength, but does not
shift in position, remaining at $0.289$ a.u. (see upper panel Fig.~\ref{fig:ex_exx_sic_power}).
 Note that the exact is obtained from solving the interacting Schr\"odinger equation and coincides
with TDDFT with the exact functional, which 
satisfies all 3 {\it Conditions}. 

We had seen in Ref.~\cite{FLSM15} that EXX fulfills {\it Condition 2} in this case, although
the resonance position is off by about $0.002$ a.u. from the exact. Here we show explicitly in the middle panel of
Fig.~\ref{fig:ex_exx_sic_power} that EXX for this dynamics also satisfies 
{\it Condition 1}; in fact even its peak strength changes in a
similar way to the exact functional (see middle panel Fig.~\ref{fig:ex_exx_sic_power}). 

SIC-LSD is shown in the lower panel of Fig.~\ref{fig:ex_exx_sic_power}:
peak-shifting signifying violation of {\it Condition 1}
is observed as a function of $\mT$. The resulting incomplete CT dynamics is evident in the inset. 
We note that the SIC-LSD peak drifts towards lower frequencies as the
charge begins to transfer in time, retracing its path as the charge
returns to the donor. That the peak tracks the instantaneous dynamics
is not unexpected, given the adiabatic nature of the approximation.
The direction of the peak shift (i.e. towards lower frequencies) appears to be 
consistent with the fact that the linear response frequency computed at the target
state, which gets partially occupied during the dynamics, is lower
($\w^f=0.237$ a.u.) than that computed at the initial state ($\w^i=0.287$ a.u., see Table \ref{tab:table}). 

\begin{figure}[th]
    \centering
    \includegraphics[width=0.5\textwidth]{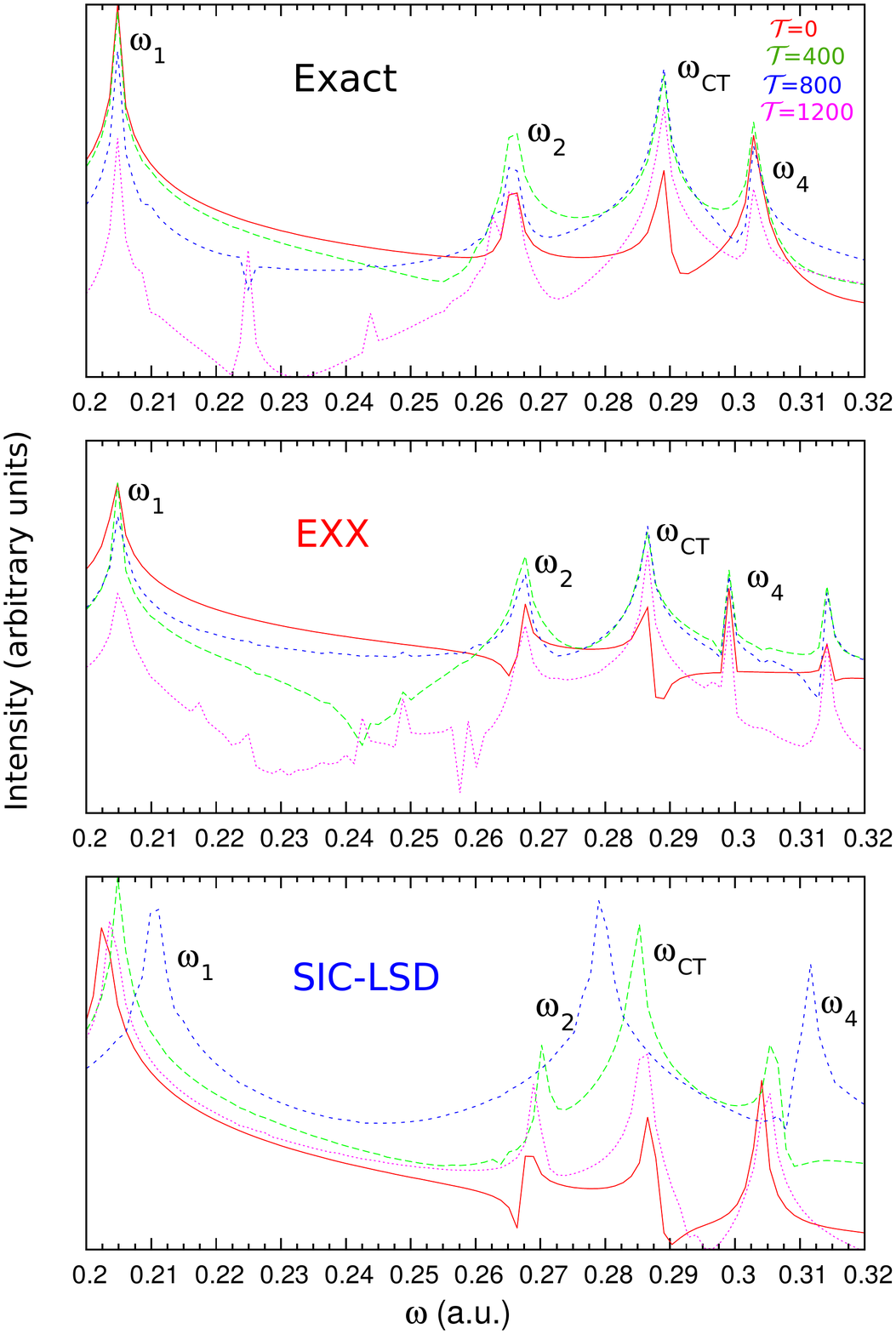}
    \caption{Logarithm of absolute spectrum $|\Delta d(\w)|$ Eq.~(\ref{eq:delta_pow_spec})
         for different laser durations $\mT$ corresponding
        to the photoexcited CT dynamics shown in insets of Fig.~\ref{fig:ex_exx_sic_power}. 
        Upper panel: Exact. Middle panel: EXX. Lower panel: SIC-LSD. 
        Exact and EXX have constant positions for all shown resonances, 
        while SIC-LSD has spuriously time-dependent resonances.  For SIC-LSD, 
        $\w_{2}$ peak shifts up in energy as the system evolves while the
        $\w_{\mathrm{CT}}$ shifts down, 
        resulting in one unique broader peak for $\mT=800$ a.u.
        (see Fig.~\ref{fig:sic_resonances} and Table.~\ref{tab:table}).}
\label{fig:real_space_broad_range}
\end{figure}

One might ask what happens to the positions of the other resonances (transitions from the photoexcited state to other unoccupied states) during the dynamics.  
In Fig.~\ref{fig:real_space_broad_range} a few peaks corresponding to excitations to higher delocalized states are analyzed as a function of $\mT$.
For the exact dynamics the positions of the peaks do not change, but as the target CT state gets populated new peaks appear in the spectra, corresponding to transitions from the CT state.
The same is true for EXX which is shown in the middle panel of
Fig.~\ref{fig:real_space_broad_range}.  This is consistent with the observation
that EXX satisfies the exact condition. As explained in Ref.~\cite{FLSM15}, this is
because once the laser is turned off, 
the KS potential for the transfering electron within EXX is constant.

\begin{figure}[ht]
    \centering
    \includegraphics[width=0.45\textwidth]{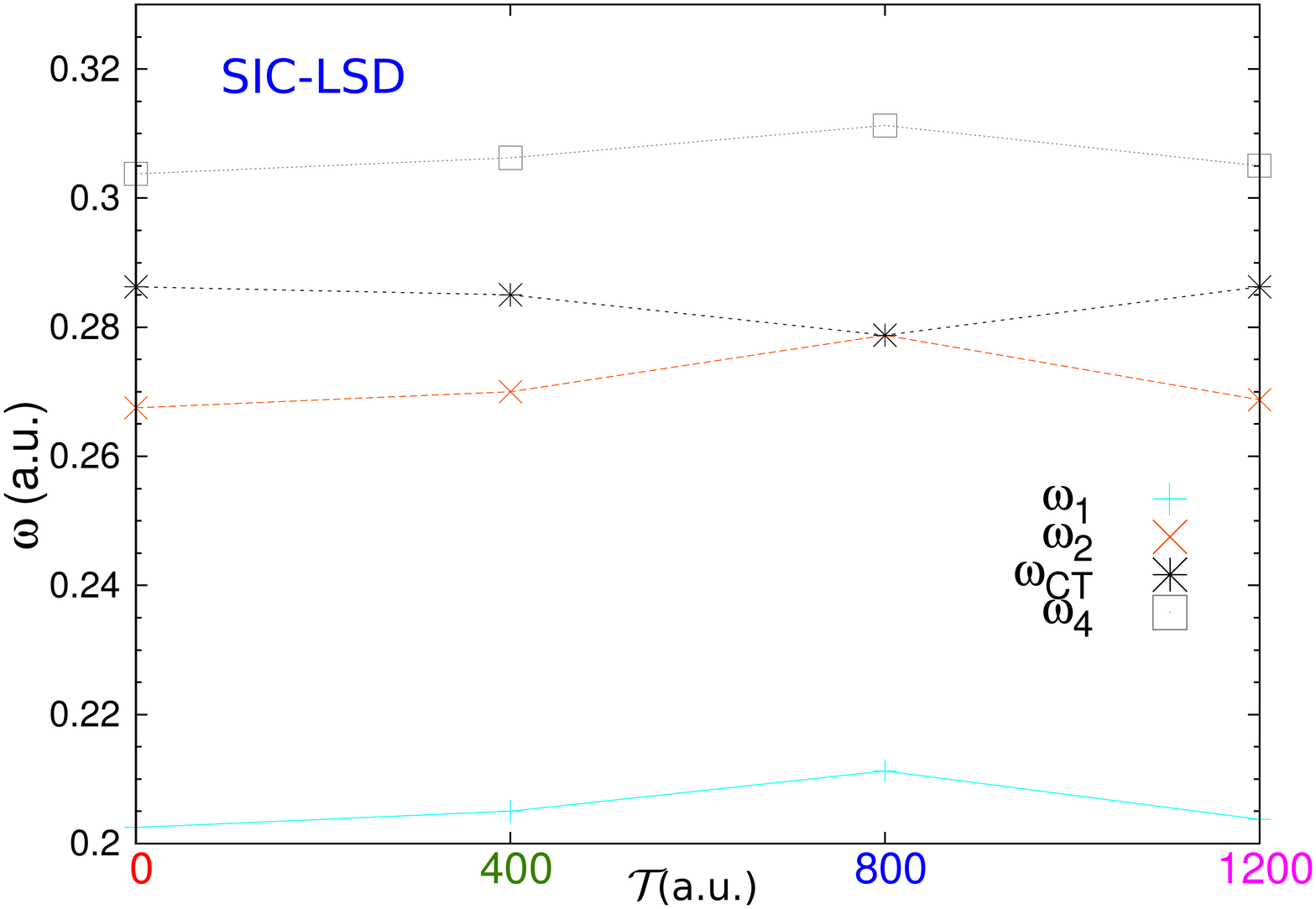}
    \caption{SIC-LSD resonances for different laser durations
        $\mT=0,400,800,1200$ a.u.. $\w_1$,  $\w_2$ and $\w_4$ shift up as charge
        transfers, but $\w_{CT}$ moves down. It can be seen in table \ref{tab:table} that the direction of the shift is towards the value of each resonance at the final target state. For $\mT=800$a.u. $\w_{2}$ and $\w_{CT}$ resonances become degenerate, resulting in the overlap of the two peaks (see lower panel Fig.~\ref{fig:real_space_broad_range}).
        All the resonances return to their initial positions as the dipole moment $d(t)$ shown in inset lower panel Fig.~\ref{fig:ex_exx_sic_power} returns to its initial value.}
\label{fig:sic_resonances}
\end{figure}

For SIC-LSD shown in lower panel of Fig.~\ref{fig:real_space_broad_range} not
only the CT resonance but also the other peaks corresponding to transitions from the initially photoexcited state to other unoccupied states shift in time. 
The direction of the shift during the transfer (from $\mT$=0 to about $\mT$= 800 au) is again towards the positions computed from the final 
target CT state (see Table \ref{tab:table} last column), and then they shift back as the charge returns, as can be seen in Table~\ref{tab:table}. 
In the table,  the SIC-LSD response frequencies computed using a $\d$-kick at different $\mT$ for a 
few transitions lying close to the CT peak are tabulated. The last column shows the values of the resonances computed applying a strong kick
in the target SCF CT state. 
The $\w_2$ transition shifts up in energy and in the target configuration it lies higher in energy than the CT resonance, 
thus these two transitions cross as the system evolves becoming degenerate at $\mT=800$ a.u.(see Fig.~\ref{fig:sic_resonances}).

These resonant frequencies can also be computed from linear response around the SCF CT state, 
by shifting the transition frequencies from this state to the relevant
unoccupied states by the value of the transition 
frequency between the CT and the locally excited state (see Table \ref{tab:table2}). 
That is, this table is a check on the "consistency" statement in {\it Condition 2}.
We have checked that the exact functional and EXX give identical numbers for the
second and third columns, but as evident in Table~\ref{tab:table2}, SIC-LSD
shows some deviation from consistency. It is interesting to note that the
frequencies obtained from the shifted linear-response TDDFT calculations in
column 3 of Table \ref{tab:table2} are close to, but not equal, to those
obtained by the second-order response calculation in the last column in Table
\ref{tab:table}. 
One can also check the consistency statement for excitations out of the CT state
(3rd excited state), by shifting the linear-response TDDFT frequencies using the
photo-excited state (4th state) as reference. This is shown in
Table~\ref{tab:table3}. It is evident that the consistency statement is more
significantly violated in this case. Consistency condition is likely to be
violated whenever {\it Condition 2} is significantly violated.  

\begin{center}
\begin{table}
\begin{tabular}{|c|c|c|c|c|c|}\hline
 SIC-LSD & $\mT=0$ & $\mT=400$ & $\mT=800$ & $\mT=1200$ & CT target state \\\hline
$\omega_1$ &0.202& 0.205 & 0.211 & 0.204 & 0.251 $\pm$ 0.003\\\hline
$\omega_2$ &0.268 & 0.270 & 0.279 & 0.269 &  0.314 $\pm$ 0.003\\\hline
$\omega_{CT}$ &0.286 & 0.285 & 0.279  & 0.286 & 0.236 $\pm$ 0.003\\\hline
$\omega_4$ & 0.304 &  0.306 & 0.311 & 0.305 & 0.352 $\pm$ 0.003\\\hline
\end{tabular}
\caption{SIC-LSD resonant frequencies (in atomic units) as computed in the
    initial locally photoexcited SCF state ($\mT=0$) and after the laser has acted
    for $\mT=400,800$ and $1200$ a.u.. 
    The values at each $\mT$ were obtained via linear response to a $\d$-kick
    perturbation of strength $0.002$ a.u.., for total propagation time $5000$ a.u. and
    resolution of $0.00125$ a.u., also shown in lower panel Fig.~\ref{fig:real_space_broad_range}.
    $\w_1$, $\w_2$ and $\w_4$ correspond to excitations from the initial photoexcited state 
    to higher, delocalized states and $\w_{\mathrm{CT}}$ is a de-excitation to the target CT state. 
    The last column shows the SIC-LSD resonances as computed from the targeted SCF CT state,
    computed from second order response using a $\d$-kick  perturbation of
    $0.1$ a.u. strength and evolving for $2000$ a.u., with a resolution of $0.003$a.u. (see text).}
\label{tab:table}
\end{table}
\end{center}
\begin{center}
\begin{table}
\begin{tabular}{|c|c|c|}\hline
 SIC-LSD & photoexc. (4th state) & CT (3th state) \\\hline
$\omega_1 = \w_{45}$ &0.202 &  $\w_{35}- \w_{34} = 0.250 $ \\\hline
$\omega_2 = \w_{46}$ & 0.268 & $\w_{36}- \w_{34} = 0.313 $  \\\hline
$\omega_{CT} = \w_{43}$ &-0.286 & $\w_{34} = 0.237  $ \\\hline
$\omega_4 = \w_{47}$ & 0.304 &  $\w_{37}- \w_{34} = 0.347 $ \\\hline
\end{tabular}
\caption{Second column: SIC-LSD resonant frequencies (in atomic units) corresponding to transitions from the
    initial locally photoexcited SCF state(which corresponds to the 4th excited state) to higher excited states, as computed from the initial photoexcited SCF state 
   (same as shown in second column Table \ref{tab:table}).
   The last column shows the SIC-LSD resonances corresponding to the same transitions but computed via linear response from the targeted SCF CT state(the 3th excited state).  (See consistency condition in Section \ref{sec:ex_cond}). The latter are obtained 
   as the difference between the transition frequencies from the SCF CT state to the same final states,
   after subtraction of the CT resonance as computed at the SCF CT state, namely $\w_{\mathrm{CT}}=\w_{34} = 0.237$ a.u.}
\label{tab:table2}
\end{table}
\end{center}
\begin{center}
\begin{table}
\begin{tabular}{|c|c|c|}\hline
 SIC-LSD & CT (3th state) & photoexc. (4th state) \\\hline
 $\w_{34}$ &0.237 & $\w_{43} = -0.286  $ \\\hline
$\w_{35}$ &0.487 &  $\w_{45}- \w_{43} = 0.489 $ \\\hline
$\w_{36}$ & 0.550 & $\w_{46}- \w_{43} = 0.554 $  \\\hline
$\w_{37}$ & 0.584 & $\w_{47}- \w_{43} = 0.590 $ \\\hline
\end{tabular}
\caption{Second column: SIC-LSD resonant frequencies (in atomic units) corresponding to transitions from the
  SCF CT state (the 3th excited state) to higher excited states, as computed from the SCF CT state.  The last column shows the SIC-LSD resonances corresponding to the 
  same transitions but computed via linear response from the initial locally photoexcited SCF state  (the 4th excited state). (See consistency condition in Section \ref{sec:ex_cond}). 
  The latter are obtained as the transition frequencies from the SCF locally excited state to the same final states, after subtraction of the CT resonance as computed at the SCF locally excited state, 
  namely $\w_{\mathrm{CT}}=\w_{43} = -0.286$ a.u.}
\label{tab:table3}
\end{table}
\end{center}
\subsection{Charge-transfer from a Ground State}
\label{sec:AE}
\begin{figure}[t]
    \centering
    \includegraphics[width=0.5\textwidth]{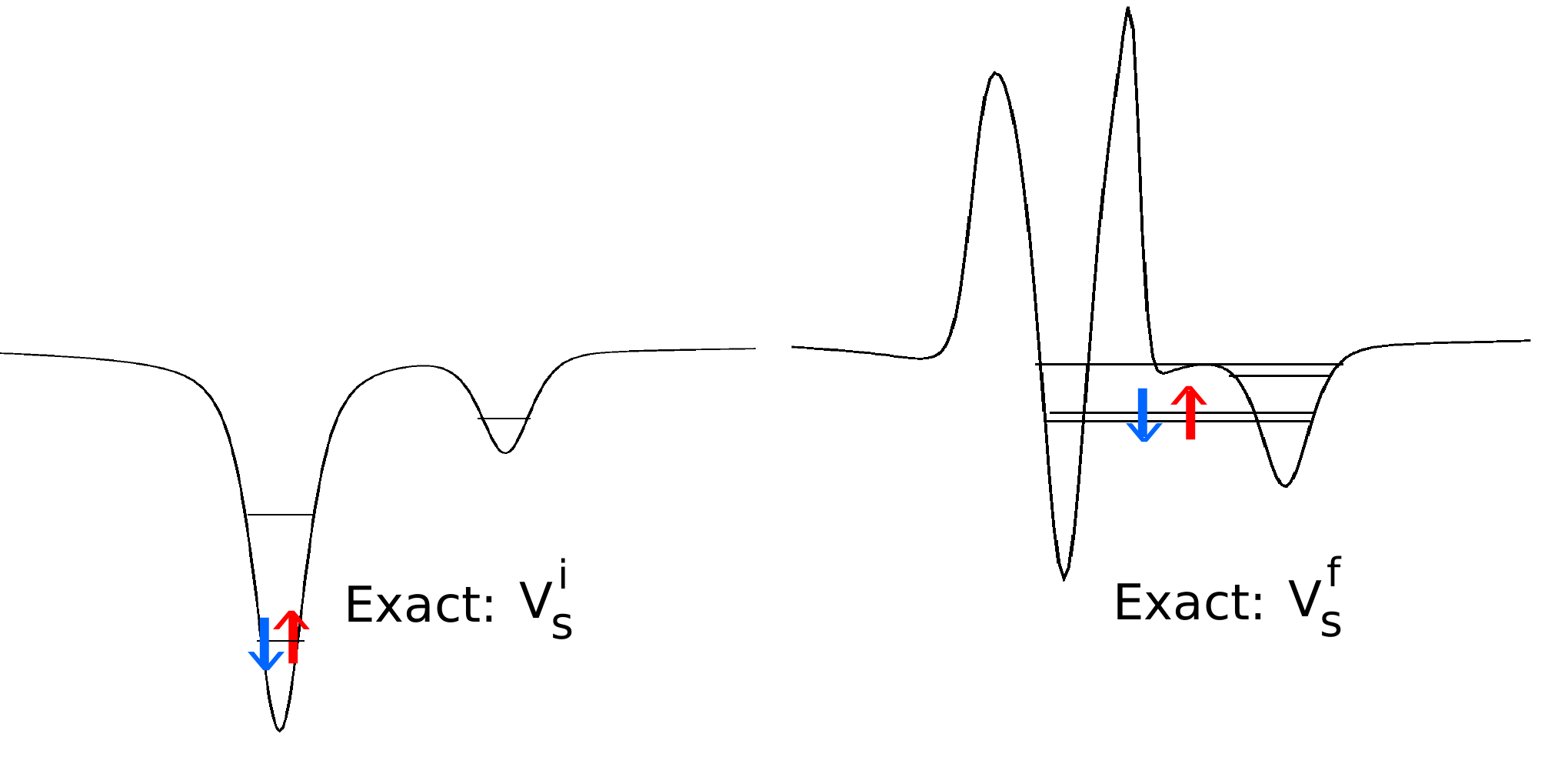}
    \caption{To the left the exact KS potential and initial KS ground state; to the right the exact final KS
    potential and target CT state are plotted. The model is specified in Section \ref{sec:photoCT}.
    As discussed in the text for the CT dynamics starting in the ground state the choice of a SSD forces the occupied time-dependent KS orbital 
    to describe both the electron that transfers to the acceptor, and the electron that remains in the  donor.  The degeneracy between the bonding and antibonding orbitals in the target CT state is related to the 
    building up over time of a CT step in the exact xc potential \cite{FERM13}.}
\label{fig:gsCT}
\end{figure}

The TDKS description of long-range CT beginning in a ground state is
quite different than that beginning in a photoexcited state.
The reason is that when beginning in a ground state the natural choice for the
KS initial state is a non-interacting ground state, which is a single-Slater determinant. Such a choice
places the transfering electron in a doubly occupied orbital, with the other electron occupying this orbital remaining in the donor.
The time-dependent orbital describing the transfering electron must
at the same time describe an electron that remains in the donor. 
The exact KS potential for a model molecule consisting of two closed-shell fragments in its ground state is depicted
in Fig.~\ref{fig:gsCT}. The doubly-occupied orbital
initially localized on the donor becomes increasingly delocalized
over both the donor and acceptor~\cite{FERM13}. In the exact
xc potential, a step feature develops over time.
Approximate functionals can not capture the step, and it is
known~\cite{FERM13,RN11,FM14,FM14b, HRCLG13} that they yield poor CT
dynamics, failing to transfer the charge, even when the
functional yields a good prediction for the CT excitation energy. This is
borne out in the dipole dynamics shown in the upper panel Fig.~3
in Ref.~\cite{RN11}, Fig.~4 in Ref.~\cite{FERM13} and upper panel Fig.~1 in
Ref.~\cite{FLSM15}.

In Ref.~\cite{FLSM15}, the failure of the approximate TDDFT dynamics was analyzed from the viewpoint of
the violation of {\it Condition 2} using a simple
two-electron model and focusing on the EXX approximation. The argument there also goes through for commonly used functionals.
Essentially, even when the CT excitation frequency, as
determined by linear response from the ground state is accurate, the
CT frequency as determined by linear response from the target CT state
is poor -- in fact it is vanishing, due to static correlation, with
delocalized antibonding type of orbitals lying near the delocalized
bonding-type of orbital. 
In the target CT state, the bare KS frequency
becomes very small, and so does the approximate $f\xc$-correction, resulting in large violations 
of {\it Condition 2} by available functionals, $\w^i>>\w^f$. 
The dynamics is consequently very poor, yielding practically no CT,
and very little dynamics occurs except at very short times.
Although {\it Condition 2} is strongly violated, the
lack of dynamics means that actually {\it Condition 1} is satisfied; since
there is very little change in the density, $v\xc(t)$ remains about
constant and so do the time-dependent spectra. This example highlights
the importance of considering both conditions together when
understanding and predicting the performance of approximate functionals.

It is also interesting to analyse {\it Condition 3} in this example. 
Let's imagine running the same CT dynamics discussed above but backwards, i.e. 
starting in the CT state and targeting the ground state. We could choose as initial state a doubly occupied orbital, in which case, 
for a stretched molecule, the bare KS eigenvalue difference $\w\s$ becomes very small due to the degeneracy between bonding and antibonding orbitals. 
The EXX kernel together with the Hartree contribution, $f\Hx$, equals half Hartree in this case, and the predicted resonance is very small and inaccurate, since the EXX kernel can not correct  the vanishing $\w\s$. 
On the other hand, we could choose as initial state a spin-broken configuration where $\uparrow$ and $\downarrow$ electrons occupy 
different orbitals as we did for the photoexcited CT dynamics studied in Section \ref{sec:photoCT}. Such choice of KS initial state 
has a finite bare KS eigenvalue difference $\w\s$, and despite vanishing of the EXX kernel correction within SPA  
as discussed in Appendix \ref{sec:spa_EXX}, EXX gives a reasonably accurate prediction of the physical 
resonance ($0.287$ au). This example illustrates how two distinct choices of KS initial state can lead to two different predictions of the 
TDDFT resonance, signifying violation of {\it Condition 3}.

\subsection{Adiabatically-Exact Propagation}
\begin{figure}[th]
 \includegraphics[width=0.45\textwidth]{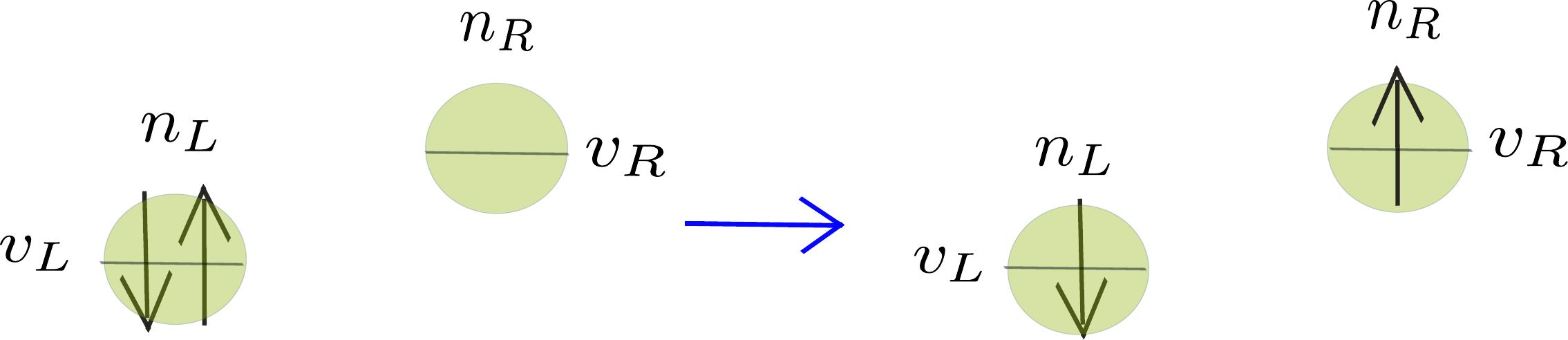}
 \caption{Two-site lattice model Eq.~(\ref{eq:HubbardH}) used to study a CT starting in the ground state. 
 Initially, the two electrons occupy the site with the deeper on-site potential $v_L$, $|\Delta n_{gs}| \approx 2$.
The target CT state consists of two open-shell fragments with approximately one electron on each, $|\Delta n_{\rm CT}| \approx 0$.}
 \label{fig:CT_gs_Hub}
 \end{figure}

To assess the impact of the adiabatic approximation itself on the fulfillment
of the {\it exact conditions}, we now consider propagating
self-consistently using the exact ground state xc
functional. This ``adiabatically-exact'' (AE)
approximation~\cite{TGK08}, $v\xc^{\mathrm{AE}}[n;\Psi_0,\Phi_0](t) = v\xc^{\mathrm{ exact-gs}}[n(t)]$, is the best possible ground state approximation to the xc functional, when using explicit density functionals.
All discrepancies with the exact dynamics will be
due to the adiabatic approximation.  Because this involves finding the
potential whose ground state density coincides with the instantaneous
one at each time-step, it is computationally quite laborious, so we
simplify the model here and consider an asymmetric Hubbard
dimer~\cite{FM14,FM14b},  beginning the dynamics in the ground state.  The exact ground state functional can be
found by Levy-Lieb contrained search within the small
Hilbert space \cite{FFTAKR13}.  In Ref.~\cite{FM14, FM14b} an asymmetric Hubbard dimer was
used to study CT dynamics from the ground state (see Fig. \ref{fig:CT_gs_Hub}), showing the same
trends as CT dynamics in the LiCN molecule~\cite{RN11}
and in the real-space one-dimensional model systems studied in Refs. \cite{FERM13, FLSM15}.
We now consider the dynamics in the light of {\it Condition 1} and {\it Condition 2}.
Further, we exploit these conditions to apply a modified
field that enhances the amount of charge transfered.

The Hamiltonian of the 2-site Hubbard model reads (Fig. \ref{fig:CT_gs_Hub}),
\ben
\begin{split}
    \hat{H}= & -T \sum_\sigma \left( \hat{c}_{L\sigma}^\dag \hat{c}_{R\sigma} +\hat{c}_{R\sigma}^\dag\hat{c}_{L\sigma} \right)
    + U \left( \hat{n}_{L \uparrow} \hat{n}_{L \downarrow} + \hat{n}_{R
    \uparrow} \hat{n}_{R \downarrow}\right) \\ 
    & +  \frac{\Delta v (t)}{2}(\hat{n}_{L} -\hat{n}_{R}),
    \label{eq:HubbardH}
\end{split}
\een
where $\hat{c}_{L(R)\sigma}^{\dag}$ and $\hat{c}_{L(R)\sigma}$ are creation and 
annihilation operators for a spin-$\sigma$ electron on the left(right) site $L(R)$, respectively, and $\hat{n}_{L(R)}=\sum_{\sigma=\uparrow, \downarrow}\hat{c}_{L(R)\sigma}^{\dag} \hat{c}_{L(R)\sigma}$ are the site-occupancy operators.
The occupation difference $\langle \hat{n}_{L} -\hat{n}_{R}\rangle =
\Delta n$ represents the dipole in this model, $d=\Delta n$, and is the
main variable~\cite{FFTAKR13}; the total number of fermions is fixed
at $N=2$. A static potential difference, $\Delta v^{0}= \sum_\sigma
(v_{L\sigma}^0 - v_{R \sigma}^0)$, renders the Hubbard dimer
asymmetric. The total external potential $\Delta v (t)$ is given by $\Delta
v(t)= \Delta v^{0}+ 2{\mathcal E(t)}$, where the last term represents a tunable electric field 
applied to induce CT between the sites.  An infinitely long-range  molecule is represented by $T/U \to 0$.
We choose here the static potential difference as $\Delta v^0=-1.5$~U,
the hopping parameter $T=0.05$~U and the on-site interaction $U=1$ (see also Ref.~\cite{FM14b}).

\begin{figure}[th]
    \centering
    \includegraphics[width=0.5\textwidth]{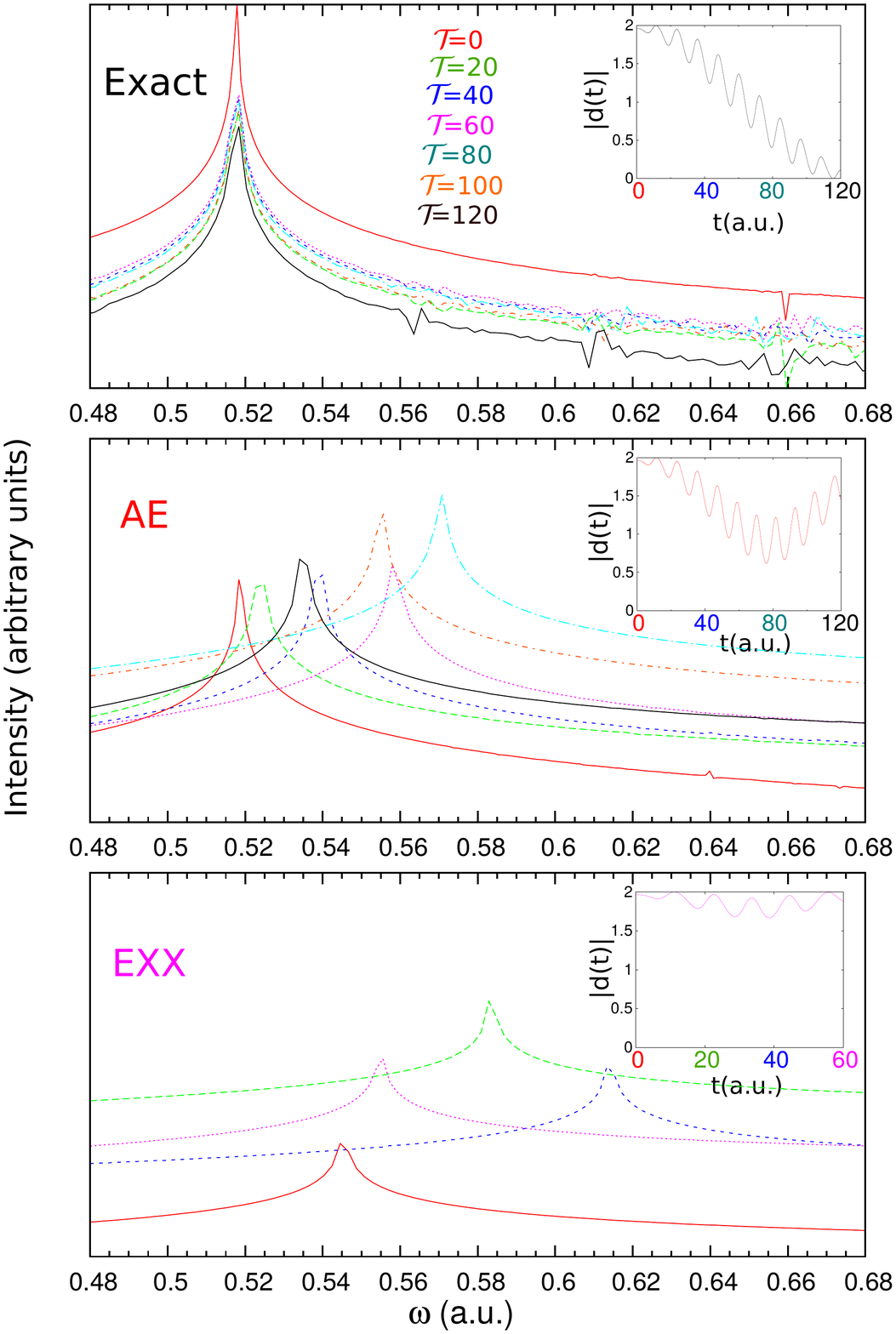}
    \caption{Logarithm  of absolute spectrum $|\Delta d(\w)|$
        Eq.~(\ref{eq:delta_pow_spec}) showing CT frequency for different laser durations $\mT$. Inset: dipole moments $|d(t)|$ for the CT dynamics starting in the ground state studied 
        in Refs.~\cite{FM14, FM14b}.  Upper panel: Exact. Middle panel: AE. Lower panel: EXX.}
\label{fig:ex_EA_EXX_shift}
\end{figure}

The KS Hamiltonian has the form of Eq.~(\ref{eq:HubbardH})
but with $U = 0$ and  $\Delta v(t)$ replaced by the KS potential difference,
\ben
\Delta v_s[\Delta n, \Phi (t_0)](t) = v\Hxc[\Delta n, \Psi (t_0), \Phi (t_0)](t) +  \Delta v(t).
\label{eq:DVs}
\een
For a resonant applied field ${\mathcal E(t)}= E_0 \sin \left(\w_{\mathrm{CT}} t \right)$,
with $\w_{CT}=(E_{\mathrm{CT}}-E_{\mathrm{gs}})=0.5177~a.u.$,  
the interacting system achieves full population of the CT state after half a Rabi
cycle (about 128 a.u.), 
coinciding with a vanishing dipole moment (see inset in upper panel of Fig.~\ref{fig:ex_EA_EXX_shift}).
AE dynamics  is poor (inset in the middle panel)
despite the AE resonance being very accurate, $\w^{AE}_{gs}=0.5187~a.u.$ \cite{FM14}.  
We follow an analogous procedure as described in Section~\ref{sec:realspace} to compute the linear response after different durations of the applied laser. The time-step used is $0.01$a.u. and the total propagation
time is $T=3000$a.u., resulting in a frequency resolution of about 
$2\pi/T \approx 0.002$ a.u..
Fig.~\ref{fig:ex_EA_EXX_shift} upper panel shows the exact dipole response $|\Delta d(\w)|$  
for different $\mT$. As expected the position of the peak is
constant at $\w_{CT}$ for all $\mT$.
In the middle panel of Fig.~\ref{fig:ex_EA_EXX_shift} the AE response for the corresponding $\mT$ is shown. 
Peak-shifting as function of the laser duration $\mT$ is noticeable, with clear violation of {\it Condition 1}.
As time evolves, the density starts transferring to the other site and
the KS potential Eq.~(\ref{eq:DVs}) follows the changes in the density. The AE
xc kernel is also time-dependent, but does not have the correct time-dependence
to maintain fixed resonance positions. 
The changes in the AE peak position follow the evolution of the density $\Delta n(t)$ 
(compare peak migration in lower panel of Fig.~\ref{fig:ex_EA_EXX_shift} with
evolution of AE dipole shown in the inset).
The AE peak shifts towards higher energies, this is consistent with our findings of Section \ref{sec:photoCT}, 
 since here $\w^f> \w^i$~\cite{FM14b} and thus the peak shift is in the direction of the resonance computed at the final state.

In the lower panel of Fig.~\ref{fig:ex_EA_EXX_shift} the EXX spectra at different
$\mT$ is shown. The EXX peak shifts as the electron starts transferring and returns 
to its initial ground state position as the density $\Delta n$ localizes back on the
donor site around $\mT=60$ a.u. (see inset lower panel of
Fig.~\ref{fig:ex_EA_EXX_shift}).
As the density transfers, the EXX peak shifts  towards higher energies,
although $ \w^{EXX}_{CT}=\w^f \to 0$.
Thus, in this case, the EXX peak does not shift towards $\w^f$.
Instead, it is consistent with the peak-shift directions observed in Ref.~\cite{PHI15}, where  the instantaneous spectra of small, closed-shell, laser-driven  molecules beginning in their ground-state, was studied.
There, a single peak was observed in the TDDFT spectra that, as the system transitions
onto a single-excited state migrates towards the value of the de-excitation energy from the doubly-excited state to the single-excited state. 

\begin{figure}[th]
    \centering
    \includegraphics[width=0.35\textwidth]{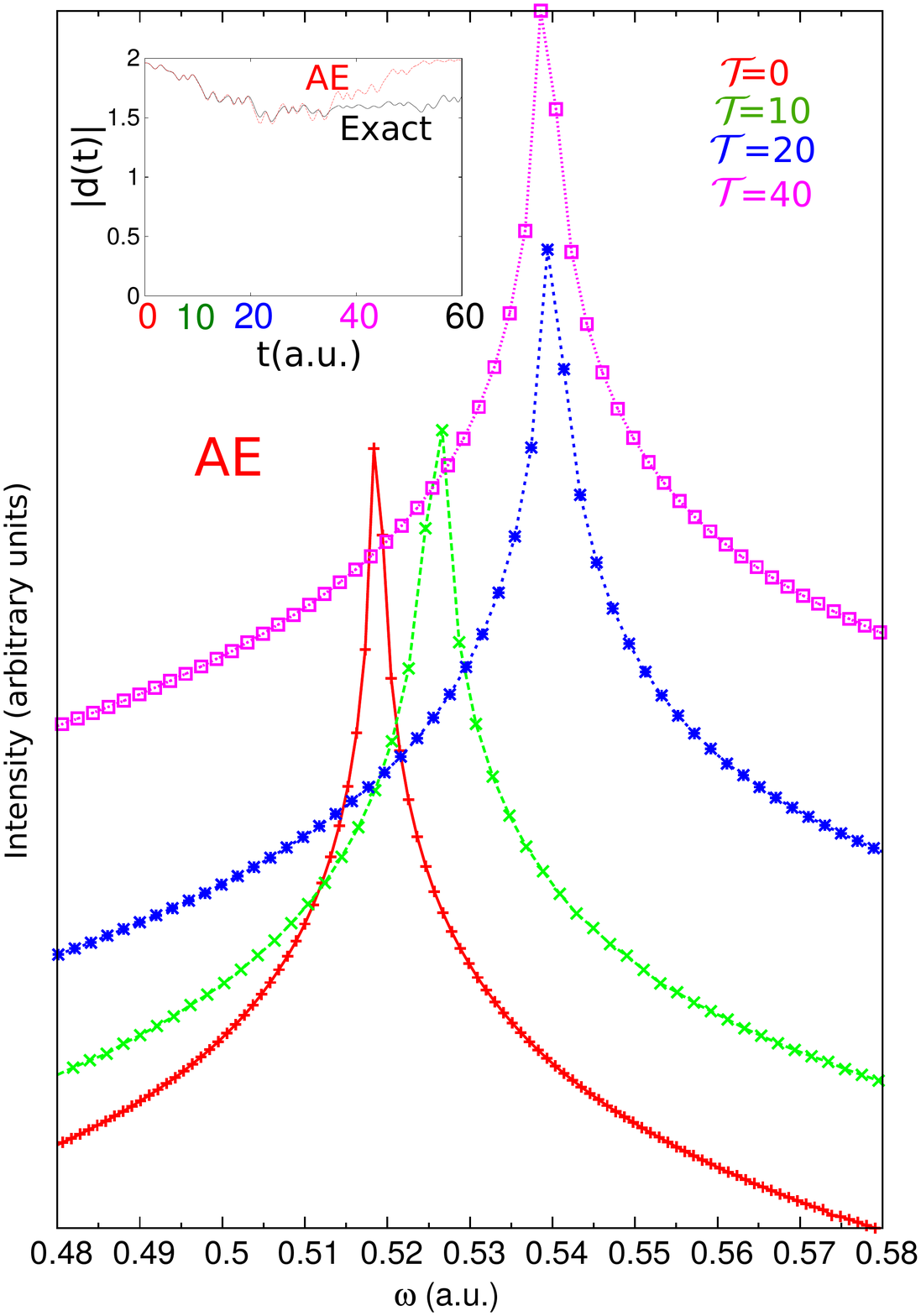}
    \caption{Logarithm of absolute AE spectrum $|\Delta d(\w)|$ showing CT resonance for different laser durations $\mT$. 
    Inset: dipole moments $|d(t)|=|\Delta n(t)|$ for exact (black) and AE (red) for an applied laser ${\cal E}(t)=1.8 \sin(0.618 t)$ starting in the ground state.}
\label{fig:AE_non_resonant}
\end{figure}

In Fig.~\ref{fig:AE_non_resonant} we present the AE response $|\Delta d(\w)|$ for {\it non-resonant} dynamics starting in the ground state.
The applied laser is detuned by $0.1$a.u. from $\w^{AE}_{gs}$ (which for this problem as discussed before is very accurate) and is chosen $20$ times stronger than the one applied in the resonant
CT dynamics of Fig.~\ref{fig:ex_EA_EXX_shift}. 
The exact dynamics is shown in the inset of Fig.~\ref{fig:AE_non_resonant} in black,
along with the AE dynamics, shown in red in the inset.
As expected for the exact dynamics, the position of the CT peak is constant and only the intensity varies (not shown here). 
Again AE presents spurious time-dependent CT resonance as a function of $\mT$, signifying violation of {\it Condition 1}.
This example of non-resonant dynamics is presented to stress the fact that spurious peak shifting within TDDFT is not exclusive to resonant dynamics.

We next consider an example that  violates {\it Condition 1} but satisfies {\it Condition 2}, unlike the previous examples: resonant dynamics for a weakly-correlated ($T=U=1$), 
homogeneous ($\Delta v^0=0$) Hubbard dimer~\cite{FFTAKR13}.
Due to symmetry of the Hamiltonian, the ground-state and first-excited densities are identical, $\Delta n_{gs}=\Delta n^{*}$.
{\it Condition 2} is satisfied within AE if we start the dynamics in the ground state and target the first local excitation, 
namely $\w^i_{AE}=\w^f_{AE}$, since the density of both stationary states is the same.
The AE dynamics however deviates from the exact as is shown in Ref.\cite{FFTAKR13} and this is due to violation of {\it Condition 1}.  
Linear response calculations at different moments $\mT$ for resonant Rabi oscillations show significant  
peak-shifting, e.g. $\w^{AE}_{gs}=2.6~a.u.$, $\w^{AE}(\mT=15)=2.3~a.u.$.

\begin{figure}[th]
    \centering
    \includegraphics[width=0.5\textwidth]{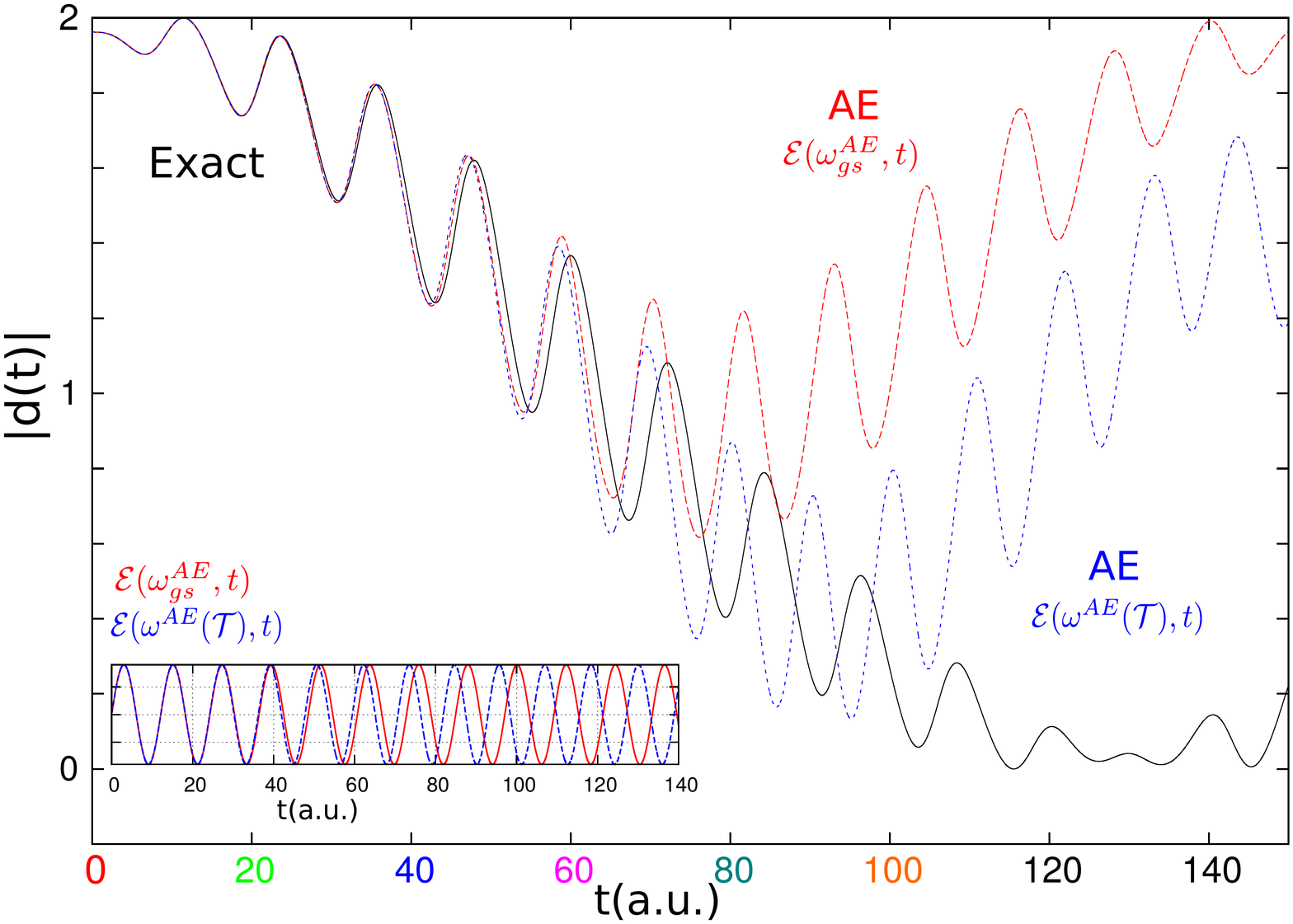}
    \caption{Exact CT dynamics starting in the ground state (black). The AE dynamics 
    in the presence of the monochromatic laser is shown in red. AE dynamics
    in the presence of the chirped laser, which every $20~$a.u. is adjusted to the instantaneous AE resonance $\w^{AE}(\mT)$ shown in middle panel Fig.~\ref{fig:ex_EA_EXX_shift}, is shown in blue. Inset: monochromatic laser
    ${\mathcal E(t)(\omega_{gs}^{AE},t)}= 0.09 \sin \left(0.518 t \right)$ in
    red, chirped laser Eq.~(\ref{eq:chirp_laser}) in blue.}
    \label{fig:HM_piecewise_laser}
\end{figure}

Finally we present a proof of principle directly demonstrating the effect of the spurious time-dependence of the electronic spectra on the TDDFT dynamics.
Peak-shifting as the system evolves means that  the instantaneous TDDFT resonance is continuously detuned from 
 the TDDFT resonance computed by perturbing  the initial state.
But what if we would adjust for this spurious detuning by making the applied laser
frequency-dependent, i.e. by designing a chirped laser that adjusts its frequency according to
 changes in the evolving density in order to stay tuned with the instantaneous TDDFT resonance of the KS system? 

In Fig.~\ref{fig:HM_piecewise_laser}, we show the results of propagating the AE system in the presence of a laser whose frequency  is adjusted piecewise-in-time during the evolution to be  approximately 
resonant with the KS system; that is every $20$a.u., the frequency of the laser is
changed such to be resonant with the
instantaneous AE CT resonances shown in middle panel of
Fig.~\ref{fig:ex_EA_EXX_shift}. 

\ben
{\cal E}[\omega^{AE}(\mT)](t)=0.09 \sin \left(\omega^{AE}(\mT)~t \right), 
\label{eq:chirp_laser}
\een
The chirped laser, Eq.~(\ref{eq:chirp_laser}), is shown in blue in the inset of Fig.~\ref{fig:HM_piecewise_laser}, 
as a guide the monochromatic ${\mathcal E(t)}= 0.09 \sin \left(0.518 t \right)$ is plotted on top in red.
In Fig.~\ref{fig:HM_piecewise_laser} the exact dipole dynamics $\Delta n(t)$ is in black, the AE dynamics under
the monochromatic laser is in red and the AE dynamics under the
chirped laser is in blue.
An improvement of the AE dynamics is observed for the laser that is  approximately ``optimally tuned'' in the way above.
Notice that for the instantaneous resonances $\w^{AE}(\mT)$ in Eq.~(\ref{eq:chirp_laser}) we have simply used the ones computed 
for the monochromatic driven AE dynamics of Fig.~\ref{fig:ex_EA_EXX_shift}.  
Further improvements and 
eventually agreement with the exact dynamics is expected if the chirped laser is designed in a self-consistent way, 
i.e. if $\w^{AE}(\mT)$ 
would be computed for the AE evolution in the presence of the chirped laser.

\section{Conclusions and Outlook}
\label{sec:conclusions}
Recent observations of time-dependent spectra in TDDFT(or TDHF) have drawn much
attention: Unphysical shifts in the position of  their spectral peaks have been reported 
for electron dynamics far from equilibrium \cite{RN12, RN12c, GBCWR13,HTPI14,FLSM15,PHI15, OMKPRV15, FCG15}.
Peak shifting is expected for coupled electron-ion dynamics or when computing resonances in the presence of time-dependent fields,
but once all time-dependent fields are turned off electronic resonance positions 
are constant for pure electron dynamics.  
The cancellation of spurious time-dependence to yield constant resonances 
has been rationalized as an exact condition for the xc functional of TDDFT
in Ref.~\cite{FLSM15}.
In this article we have elaborated on the theoretical aspects of 
the formulation and provided a detailed study of some representative examples 
that illustrate the relevance of the exact condition for the dynamics.

A generalized non-equilibrium response function was derived which applies
to general non-equilibrium situations when the external fields are off \cite{FLSM15, PS15}. 
In contrast to the standard linear response formalism applied to systems around the ground state, the 
generalized non-equilibrium response function $\tilde\chi[n^{(0)}_\mT; \Psi(\mT)]$ deals with non-stationary densities 
$n^{(0)}_\mT$ in the absence of time-dependent fields and is not time-translationally invariant. We showed that due to the lack of this symmetry the frequency-dependent response 
$\tilde{\chi}_{t'}(\br, \br',\omega)$ or $\tilde{\chi}_{T}(\br, \br',\omega)$ depends parametrically on a time-variable $t'$  or $T=(t+t')/2$, respectively, and each 
exhibits a different pole structure. The density response $\delta n(\br,\w)$ is independent of this choice as it should be. 
The latter has poles at the physical resonances of the system, corresponding 
to transitions between eigenstates of the unperturbed Hamiltonian.

By virtue of the Runge-Gross theorem  \cite{RG84} the exact time-dependent KS system reproduces the non-equilibrium density response $\delta n (\br,\w)$
exactly. Therefore, a Dyson-like equation connects $\tilde\chi[n^{(0)}_\mT; \Psi(\mT)]$ with the non-equilibrium KS response 
$\tilde\chi\s[n^{(0)}_\mT; \Psi(\mT), \Phi(\mT)]$ via a generalized kernel  $\tilde{f}\Hxc[n^{(0)}_\mT; \Psi(\mT), \Phi(\mT)]$. 
A simple Lehmann representation can be written down for $\tilde\chi$ but not for $\tilde\chi\s$ because 
the KS potential is not static for non-stationary densities  $n^{(0)}_\mT$.

The exact condition was formulated in several different ways. In the
language of pump-probe experiments, the resonance positions must be
independent of the moment $\mT'$ the pump is turned off and of the
delay $\theta$ between pump and probe. More generally, in the absence of
ionic motion, the TDDFT response frequencies (corresponding to transitions between two given interacting states) of a field-free system
must be independent of the interacting state $\Psi(\mT)$ around
which the response is calculated ({\it Condition 1}).  When the
interacting system is in any excited stationary state, and the KS reference state is also chosen to be a stationary state of its potential, we can formulate
the exact condition in a matrix form. In such a case, 
the TDDFT resonance frequency for a specific
transition between two stationary states must be independent of which of the two states the response is calculated around ({\it Condition 2}).  Further, a consistency condition relates the frequencies of transitions to other states from each reference state. 
%Even more, 
%all transition frequencies must be identical regardless of the state of referen%ce, which implies a consistency condition. 
The TDDFT response
frequencies must be invariant
also with respect to the choice of KS state ({\it Condition 3}).  
The exact conditions pose a very challenging task for
the xc kernel, as we illustrated with several examples.

The first  example was that of CT dynamics
from a photo-excited state. We used a model system of two electrons in an asymmetric double-well and started the KS propagation with 
one orbital  promoted to a locally excited orbital; this electron was then driven to the 
other well by means of a weak resonant field. 
The performance of the different functionals to accurately simulate this dynamics was related to the fulfillment of the exact condition
in its different forms.
In Ref.~\cite{FLSM15} it was observed that the degree of violation 
of {\it Condition 2} was directly related to the performance of the approximate functional to reproduce the dynamics.
Here the time-dependent spectra for the same approximated functionals, 
including EXX and SIC-LSD are compared against
the exact calculation at different moments in the evolution. We find that in addition to {\it Condition 2} also {\it Condition 1} 
was violated within SIC-LSD, resulting in spurious peak shifting and consequently in poor dynamics. 
The CT peak moved towards the 
SIC-LSD resonance computed around the final target state and this trend held also for the other resonances of the system. SIC-LSD resonances were also shown to violate the consistency condition. 
%(see Tables \ref{tab:table2} and \ref{tab:table3}).
Unrestricted EXX was shown  to fulfill {\it Condition 2} for this particular case~\cite{FLSM15} and here we showed it also has fixed peak positions along 
the evolution (fulfillment of {\it Condition 1}), resulting in very accurate photoexcited CT dynamics.

In the next example, CT dynamics starting in the ground state, EXX
trivially fulfilled {\it Condition 1} but violated {\it Condition 2},
resulting in poor dynamics. We also briefly discussed the violation of
{\it Condition 3} by EXX when we consider different initial KS states
when running the dynamics backwards beginning in the CT state.

In order to assess the impact of the adiabatic approximation alone,
independent of the effect of the choice of approximate ground-state
functional, we used a two-site lattice model.  Given the small Hilbert
space the exact ground state xc functional can be found
\cite{FFTAKR13} and was used to self-consistently propagate the KS
system (AE propagation). We tuned the parameters of the system to mimic
long-range CT dynamics starting in the ground state~\cite{FM14,
  FM14b}.  We observed that the AE CT dynamics violates both {\it
  Condition 1} and {\it Condition 2} and resulted in poor dynamics.
Violation of {\it Condition 1} was observed for AE for both resonant
and also off-resonant dynamics. In the case of the weakly-correlated
homogeneous Hubbard model studied in Ref.~\cite{FFTAKR13} {\it
  Condition 2} was satisfied but {\it Condition 1} was violated within
AE, resulting in inaccurate dynamics.

Perhaps the clearest impact of the influence of the exact condition on the dynamics was illustrated
by the ``optimally tuned" laser that adjusted in a piece-wise manner to the instantaneous AE resonance (Figure~\ref{fig:HM_piecewise_laser} and discussion). This showed that the AE propagation 
in the presence of this chirped laser resulted in an improved charge transfer rate in the case of resonant CT dynamics.

We conclude that in order to be able to predict the performance of a given approximate functional all three {\it Conditions}
need to be considered.
Further because the best possible ground state approximation for the density-functional fails to fulfill the exact conditions, 
resulting in poor dynamics, we stress the need to go beyond the adiabatic approximation.

We have shown the large impact that the violation of the exact
conditions has on the ability of approximate functionals to reproduce
the dynamics: the higher the degree of the violation, the poorer the
simulated dynamics. The examples presented here are drastic, but we
have only analyzed small systems, perhaps a worst-case scenario for
approximate TDDFT.  An important question for future work is the
system-size scaling of the violation of the exact conditions.  When
ionic motion is considered, the spectral peaks can be vibrationally
broadened, perhaps relaxing the stringent exact conditions discussed
here.  Development of an approximate functional or a propagation
scheme that fulfills exactly or approximately the exact conditions is
an important direction for future research.

\begin{acknowledgments} 
    Financial support from the National Science Foundation CHE-1152784 (for K.L.),
    Department of Energy, Office of Basic Energy Sciences, Division of Chemical
    Sciences, Geosciences and Biosciences under award DE-SC0015344(N.T.M., J.I.F.),
    a grant of computer time from the Cuny High Performance Computing Center under
    NSF grants CNS-0855217 and CNS-0958379. 
\end{acknowledgments}

\bibliographystyle{apsrev4-1}
%\bibliography{ref}
\appendix
\section{$f\Hxc$ correction in EXX within the Single Pole Approximation}
\label{sec:spa_EXX}
Within the single-pole approximation using EXX it can be shown that for a
two-electron singlet state where the two electrons occupy different spatial orbitals, the $f\Hxc$ correction vanishes.
First, note the definition of the spin-resolved xc kernel~\cite{PGG96,GPG00},
\ben
f\xc^{\sigma\sigma'}(\br,\br',t,t') = \frac{\d
  v\xc^{\sigma}(\br,t)}{\d n_{\sigma'}(\br',t')} 
\label{eq:fxcspindec}
\een 
In adiabatic EXX, 
$v\xc^{\sigma}[n_\uparrow,n_\downarrow](\br) = v\x^\sigma[n_\uparrow,n_\downarrow](\br) =\frac{\d
  E\x[n_\uparrow,n_\downarrow]}{\d n_\sigma (\br)}$,  where for two electrons
\bea E\x &=& -\half\sum_{\sigma=\uparrow,\downarrow}
\sum_{ij}^{N_\sigma} \int\!\!\int \frac{\phi_{i\sigma}(\br)
  \phi_{i\sigma}(\br') \phi_{j\sigma}(\br') \phi_{j\sigma}(\br)}{|\br
  - \br'|} \dr\dr' \nonumber \\ &=&
-\half\sum_{\sigma=\uparrow,\downarrow} \int\!\!\int
\frac{n_{\sigma}(\br) n_{\sigma}(\br')}{|\br - \br'|} \dr\dr'
\nonumber \\ &=& 
- E\H[n_\downarrow] - E\H[n_\uparrow]\;. \eea
This yields
$v\x^\sigma[n_\uparrow,n_\downarrow] =-v\H[n_\sigma]$, and so using Eq.~\ref{eq:fxcspindec},
\ben
    f\xc^{\downarrow \uparrow} = 
     f\xc^{\uparrow \downarrow} = 0 \;\; {\rm while}\;\; f\xc^{\uparrow \uparrow} = f\xc^{\downarrow \downarrow} =  -f\H 
\een
Thus for the parallel spin case 
the Hartree-exchange kernel $f\Hx^{\sigma \sigma}=0$ vanishes, however in the case of
anti-parallel spin $f\Hx^{\uparrow \downarrow}=  f\Hx^{\downarrow \uparrow}=f\H$. 

Now for a non-degenerate Kohn-Sham pole, $\omega\s = \e_a - \e_i$,
the single-pole approximation for the TDDFT excitation energy depends only on
the parallel-spin component of the kernel~\cite{PGG96,GPG00}:
\ben
\omega  = \omega\s + \int \Phi_q(\br)\left(f\H + f\xc^{\sigma\sigma}\right)\Phi_q(\br') d\br d\br'
\label{eq:spa-nondeg}
\een
(where, as before, $\Phi_q(\br) = \phi_i(\br)\phi_a(\br)$), clearly yielding a zero correction for the case of adiabatic EXX. 
That is, within the single-pole approximation, the KS response frequencies are
identical to the TDDFT excitations when the two electrons are occupying
different orbitals (see Section \ref{sec:photoCT}). In contrast, when they occupy the same spatial orbital, as for a
spin-saturated two-electron singlet (see section \ref{sec:AE}), each Kohn-Sham pole is doubly-degenerate,
and Eq.~\ref{eq:spa-nondeg} is modified, see Ref.~\cite{PGG96,GPG00} for details; the kernel correction in EXX, $f\x=-f\H/2$, does not vanish for this case. 
\end{document}